\documentclass[aps,pre,longbibliography,twocolumn,superscriptaddress,showpacs]{revtex4-1}

\usepackage{CJK}
\usepackage{graphicx}
\usepackage{amsmath}
\usepackage{dcolumn}
\usepackage{bm}
\usepackage[normalem]{ulem}
\usepackage{xcolor}
\usepackage{hyperref}

\usepackage{enumitem,kantlipsum}
\numberwithin{equation}{section}
 
\usepackage{epstopdf}
\newcommand{\rf}{\rho_{ref}}

\newcommand{\bfr}{\mathbf{f}_\rho}
\newcommand{\ms}{M_{\rm spinodal}}
\newcommand{\mb}{M_{\rm binodal}}

\newcommand{\fl}{active polar ordered fluids }
\newcommand{\dr}{\delta\rho}

\newcommand{\rc}{\rho_{c}}

\newcommand{\mv}{\langle\bv\rangle}

\newcommand{\order}{\mathcal{O}}
 \newcommand{\nn}{\nonumber}

\newcommand{\bew}{\begin{widetext}}
\newcommand{\ew}{\end{widetext}}

\newcommand{\bp}{\mathbf{p}}
\newcommand{\bq}{\mathbf{q}}
\newcommand{\bQ}{\mathbf{Q}}
\newcommand{\bv}{\mathbf{v}}

\newcommand{\br}{\mathbf{r}}
\newcommand{\alp}{\alpha(P_x, \varpi)}
\newcommand{\sigp}{\sigma(P_x, \varpi)}

\newcommand{\brp}{\mathbf{r}_\perp}

\newcommand{\bqp}{\mathbf{q}_\perp}

\newcommand{\bpp}{\mathbf{p}_{_\perp}}
\newcommand{\bvp}{\mathbf{v}_\perp}
\newcommand{\bR}{\mathbf{R}}
\newcommand{\bff}{\mathbf{f}}

\newcommand{\hx}{\hat{x}}

\newcommand{\bk}{\mathbf{k}}

\newcommand{\sep}{ \ \ \ , \ \ \ }

\newcommand{\beq}{\begin{equation}}
\newcommand{\eeq}{\end{equation}}
\newcommand{\beqn}{\begin{eqnarray}}
\newcommand{\eeqn}{\end{eqnarray}}
\newcommand{\pp}{\partial}
\newcommand{\dd}{{\rm d}}

\newcommand{\dit}{I_2(\varTheta, \varXi, \varUpsilon)}
\newcommand{\dio}{I_1(\varTheta, \varXi, \varUpsilon)}

\def\rf#1{(\ref{#1})}

\newcommand{\partialder}[2]{\frac{\partial #1}{\partial #2}}

\begin{document}
\title{ Phase separation in ordered polar active fluids: A new Universality class}
\author{Maxx Miller}
\email{maxxm@uoregon.edu}
\affiliation{Department of Physics and Institute for Fundamental
 Science, University of Oregon, Eugene, OR $97403$}
\author{John Toner}
\email{jjt@uoregon.edu}
\affiliation{Department of Physics and Institute for Fundamental
 Science, University of Oregon, Eugene, OR $97403$}

\begin{abstract}
We show that phase separation in ordered polar active fluids belongs to a new universality class.
This describes large collections of self-propelled entities (``flocks"), all spontaneously moving in the same direction,  in which attractive interactions (which can be caused by, e.g., autochemotaxis) cause phase separation: the system spontaneously separates into a high density band and a low density band, moving parallel to each other,  and to the direction of mean flock motion, at different speeds.
 The upper critical dimension for this transition is $d_c=5$, in contrast to the well-known $d_c=4$ of equilibrium phase separation. We obtain the large-distance, long-time scaling laws of the velocity and density fluctuations,  which are characterized by universal critical correlation length and order parameter exponents  $\nu_\perp$, $\nu_\parallel$  and $\beta$ respectively. We calculate these to $\order (\epsilon)$ in a $d=5-\epsilon$ expansion.
\end{abstract}
\maketitle

\section{Introduction}
One of the most important ideas in Condensed Matter Physics is the concept of ``universality", which asserts that it is {\it only} the symmetries and conservation laws describing a given phase of matter, or the transitions between different states, that determine the long-distance, long-time properties of those phases and transitions\cite{chaikin, Ma}. The microscopic details of the system  in question do not affect these long-distance, long-time  properties. 

More recently, it has been realized that non-equilibrium systems and phase transitions can belong to {\it different} universality classes than their equilibrium counterparts.
A dramatic demonstration of this difference is provided by the phenomenon of ``flocking", in which a large collection of self-propelled entities, which could be macroscopic living creatures\cite{Locust}, microorganisms, or even intra-cellular components\cite{TT5}, spontaneously all move in the same direction. Synthetic examples of such ``flocks"  also abound\cite{quinke}. 

A more technical term for such a ``flock" is a ``polar ordered active fluid": ``polar" because a particular direction is picked out (namely, the direction of the mean flock velocity vector $<\bv>$), ``ordered" because this direction is the same throughout an arbitrarily large flock (i.e., the flock has ``long-ranged order"), ``active" because the ``boids" are self-propelled, which consumes energy locally, and ``fluid", because we assume that translational symmetry is {\it not} broken: we are considering flying {\it fluids}, not flying crystals. 

We consider flocks {\it without} momentum conservation (i.e., "dry" flocks). As a result, the only conservation law in ours system is  boid number: boids are not being born and dying ``on the wing''. Both ``Malthusian" flocks\cite{malthus}, in which boid number is {\it not} conserved, and ``wet'' flocks\cite{Wet1,wet2}, exhibit very different hydrodynamic behavior, which we will not discuss further here. 

The underlying symmetries of the dynamics of flocking are the same as those of ferromagnetism: rotation invariance. Likewise, the nature of the symmetry {\it breaking} is the same: by spontaneously  choosing a direction to move, a flock is breaking the underlying rotation invariance of the dynamics, in precisely the same way that a ferromagnet spontaneously breaks the underlying rotation invariance of the spin dynamics.

Despite these similarities, 
the fundamentally non-equilibrium nature of flocking makes it very different from ferromagnetism. In particular, flocks can spontaneously break rotation invariance even in  spatial
dimension $d=2$ \cite{Vicsek,TT1}, and exhibits ``anomalous hydrodynamics" \cite{TT1,TT3,TT4} even in spatial dimensions $d>2$. By ``anomalous hydrodynamics", we mean that the long-wavelength, long-time behavior of these systems can {\it not} be accurately described by a linear theory; instead, non-linear interactions between fluctuations must be taken into account, even to get the correct scaling laws. Indeed, it is the anomalous hydrodynamics in $d=2$ that makes the existence of long-ranged order possible \cite{TT1,TT3,TT4}.

\begin{figure}
    \centering
    \includegraphics[width=0.85\linewidth]{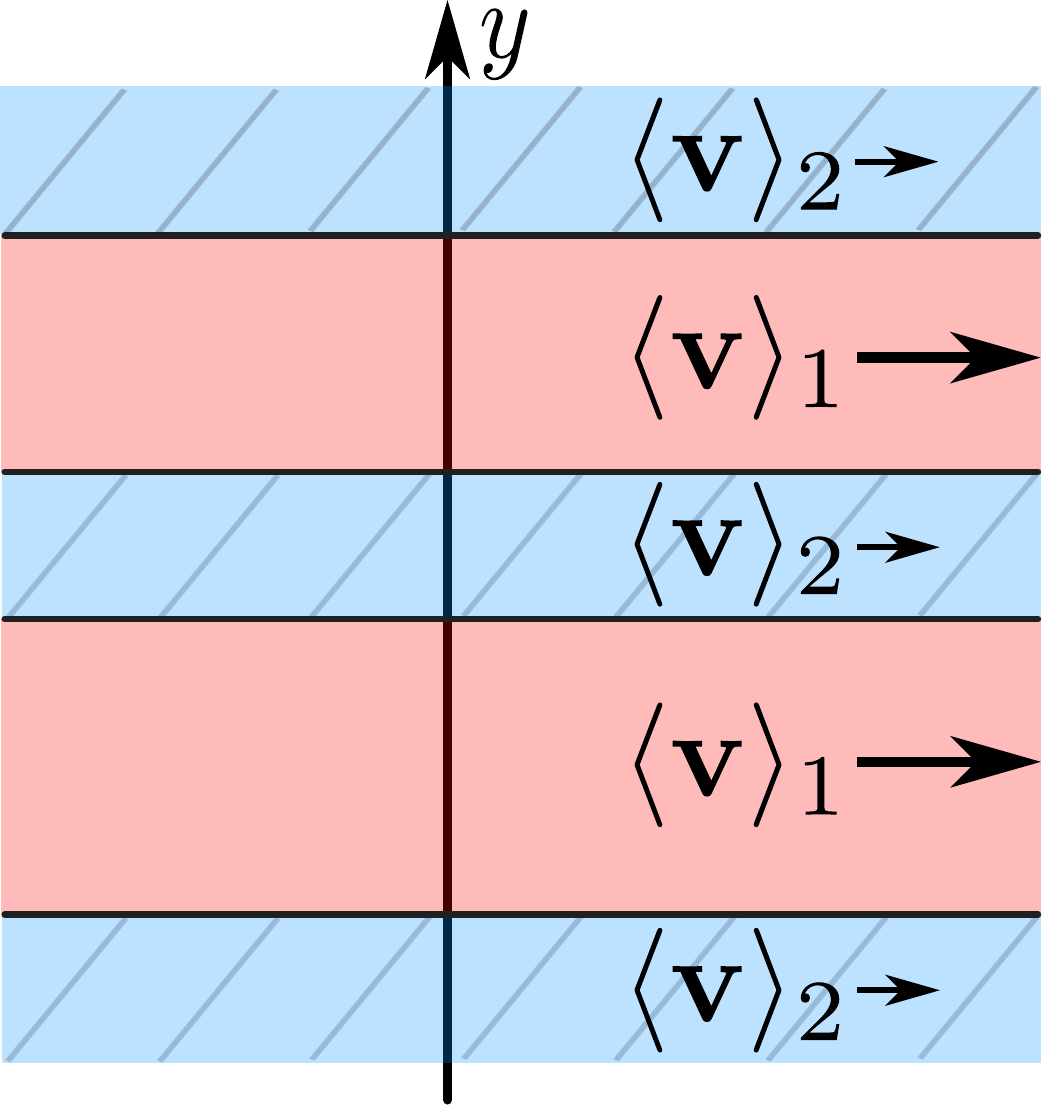}
    \caption{The ``band" structure of the instability at intermediate times. The density is only modulated along one of the directions (which we call $y$, and which is indicated in the figure) perpendicular to the direction $\hx$ of mean flock motion.}
    \label{bandsCartoon}
\end{figure}

Recently\cite{usprl, uslong} it has been realized that flocks can also exhibit another phenomenon familiar from equilibrium physics: phase separation. This occurs when the individual components of the flock attract each other, and is characterized by the separation of a large flock into one high density band, and one low density band, both moving parallel to each other at different speeds, as illustrated in figure \ref{bandsCartoon}.

One mechanism for such attraction is ``autochemotaxis", in which every member of the flock  (which we'll call ``boids") emits a   ``chemo-attractant";  i.e., a substance to which they themselves are attracted. The chemo-attractant then diffuses, and decays with a finite lifetime. The boids ``flock" - that is, follow their neighbors,  but with a bias in the direction of the gradient of the chemo-attractant concentration.

Of course, many other mechanisms besides autochemotaxis can generate attractions between the boids. Any such mechanism could lead to the phase separation in a polar ordered active fluid considered in \cite{usprl, uslong} and here.

\begin{figure}
        \centering
       \includegraphics[width=\linewidth]{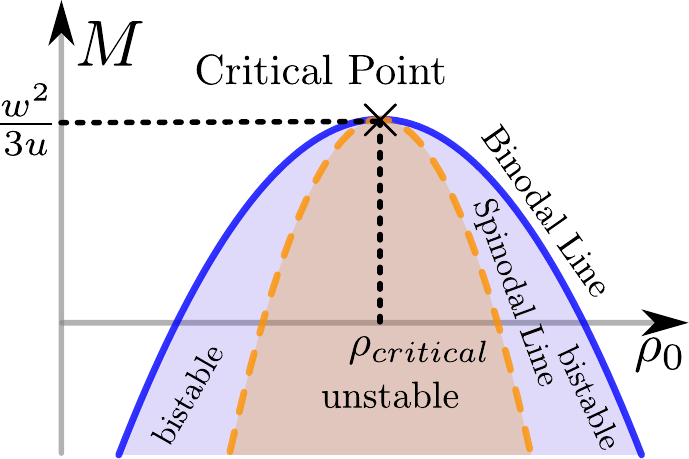}
        \caption{Phase diagram of flock phase separation. The solid blue and  dashed orange curves are the binodal and spinodal lines respectively. Note that those expressions are only valid close to the critical point. In the orange filled region  under the spinodal line, which is labeled ``unstable", only the two-phase state is stable. In the blue region between the spinodal and binodal lines, both the two phase state, and the homogeneous, one phase state, are stable. The analysis we perform in this paper is done close to the critical point, where our assumption that density variations are small is valid.}
        \label{spinodal}
\end{figure}

The treatment of this phase separation in \cite{usprl, uslong} revealed a phase diagram qualitatively identical to that found for equilibrium phase separation, as illustrated in figure \ref{spinodal}. Here the vertical axis $M$ could be any experimentally tunable parameter that decreases with increasing strength of the attractive interactions. Increasing the strength of the autochemotaxis in an autochemotactic system, for example,  which could be accomplished by increasing the strength of the boids' response to the chemical signal, or its emission rate, or the chemo-attractant lifetime, would accomplish this.

The horizontal axis is the  number density of boids per unit volume (or area, in a two dimensional system). 

This ``phase diagram" ( figure \ref{spinodal}) is to be interpreted as follows. For all $M$ above the ``binodal" parabola, which is the upper parabola in \ref{spinodal}, the system can only be in the uniform state, which is stable. For values of $M$ between the two parabolae (the blue and orange  curves in figure \ref{spinodal}) - that is, for $\mb>M>\ms$
both the two phase state, and the homogeneous, one phase state, are stable.
Finally, for $M<\ms$, only the phase separated state is stable.

The strong similarity between these results and equilibrium liquid-vapor phase separation is apparent from this phase diagram, which is identical to that for an equilibrium liquid-vapor system, with $M$ playing the role of temperature. Recall that $M$ can be increased (decreased) experimentally by decreasing (increasing) the strength of the chemotaxis, or, more generally, by tuning the strength of whatever attractive interactions in the flock reduce the inverse compressibility. 

Note there is an important difference between these results and 
equilibrium liquid-vapor phase separation. In {\it equilibrium} 
systems, a homogeneous density state is only  {\it meta}-stable in the region between the binodal 
and spinodal curves. The homogeneous state exists at a {\it local} minimum in 
the free energy; the {\it global} minimum  is the phase separated state. In contrast,  since flocks are {\it non-equilibrium} systems, there is no criterion that we know of analogous to the equilibrium global minimization principle to decide which of the two locally stable states is  ``preferred". Instead,  all we can  determine is the {\it local} stability of each 
phase between the binodal and spinodal curves. Between these two curves, both states 
are locally stable. We  therefore refer to this region as the bistable region.

The treatment of this phase separation in \cite{usprl, uslong} is entirely ``mean-field": that is, it ignores fluctuations in the local density and velocity. Fluctuations in the density are well-known\cite{chaikin, Ma} to radically change the scaling behavior of the density near the critical point in equilibrium systems. In active polar ordered flocks, fluctuations are even more important, because, in addition to the density field, which has large fluctuations because it is becoming ``soft" near (and at) the critical point, the local velocity field of the flock (or, more precisely, its components perpendicular to the mean velocity $\mv$) are Goldstone modes, and so have large fluctuations themselves. 

Hence, to understand the true behavior of the system near the critical point in figure \ref{spinodal}, and the shape of the phase boundaries themselves there, we therefore clearly must include the effect of fluctuations. We do so in this paper, by performing a dynamical RG analysis of the hydrodynamic theory of \fl  near the critical point.

We find that this critical point belongs to a completely different universality class than the equilibrium liquid-vapor critical point. Indeed, even the {\it upper critical dimension} $d_{LC}$ of the \fl critical point, defined in the usual way as the spatial dimension below which fluctuations change the scaling behavior of the transition,   is different: it is $d_{LC}=5$. In contrast, for equilibrium phase separation, the upper critical dimension is $d_{LC}=4$.

We have calculated a number of universal exponents characterizing the scaling behavior near the critical point using the dynamical RG in an $\epsilon=5-d$ expansion.  The first of these is the usual exponent $\beta$ giving the width $\dr$ of the binodal and spinodal curves in figure \ref{spinodal}. Those widths both scale as a power law in the distance $M_c-M$ from the critical point:
\begin{align}
       \delta\rho &\propto |M-M_C|^{\beta} \,.
    \label{betadef}
\end{align}
We find
\begin{align}
    \beta 
 =  \frac{1}{2} - \frac{\epsilon}{6} + \order(\epsilon^2) \,.
 \label{betaepsintro}
    \end{align}
In addition, we have calculated the correlation length exponent $\nu$. Or, to be more precise, we find {\it two} correlation length exponents $\nu_{\perp, \parallel}$ for the divergences of the correlation lengths    $\xi _{\perp,\parallel}$  perpendicular and parallel to the direction of mean flock motion, respectively. 
These are defined by
\begin{align}
    \xi _{\perp,\parallel}\propto |M-M_C|^{\nu_{\perp, \parallel}} \,.
    \label{nudef}
    \end{align}
We find
\begin{align}
    \nu_\perp = \frac{1}{2} + \frac{\epsilon}{12} + \order(\epsilon^2) \,,
    \label{nuepsintro} 
\end{align}
and 
\begin{align}
    \nu_\parallel = 1 + \frac{\epsilon}{6} + \order(\epsilon^2) \,.
    \label{nuparepsintro} 
\end{align}
In addition to this anisotropy in correlation lengths, which does not occur for equilibrium phase separation, the interpretation of the correlation lengths in phase separating flocks is also different. 
In equilibrium phase separation, the correlation length is the length scale on which density correlations decay exponentially; that is

\begin{align}
    C_{\rho\rho}(\br) \equiv \langle \delta\rho(\br +\bR)\delta\rho(\bR)\rangle \propto e^{-r/\xi} \label{rho corr eq}
\end{align}
where $\delta\rho(\br,t)\equiv \rho(\br,t)-\rho_c$ is the departure of the local number density of flockers  $\rho(\br,t)$ at position $\br$ and time $t$ from its mean value $\rho_c$. 

\begin{figure}
    \centering
    \includegraphics[width=0.8\linewidth]{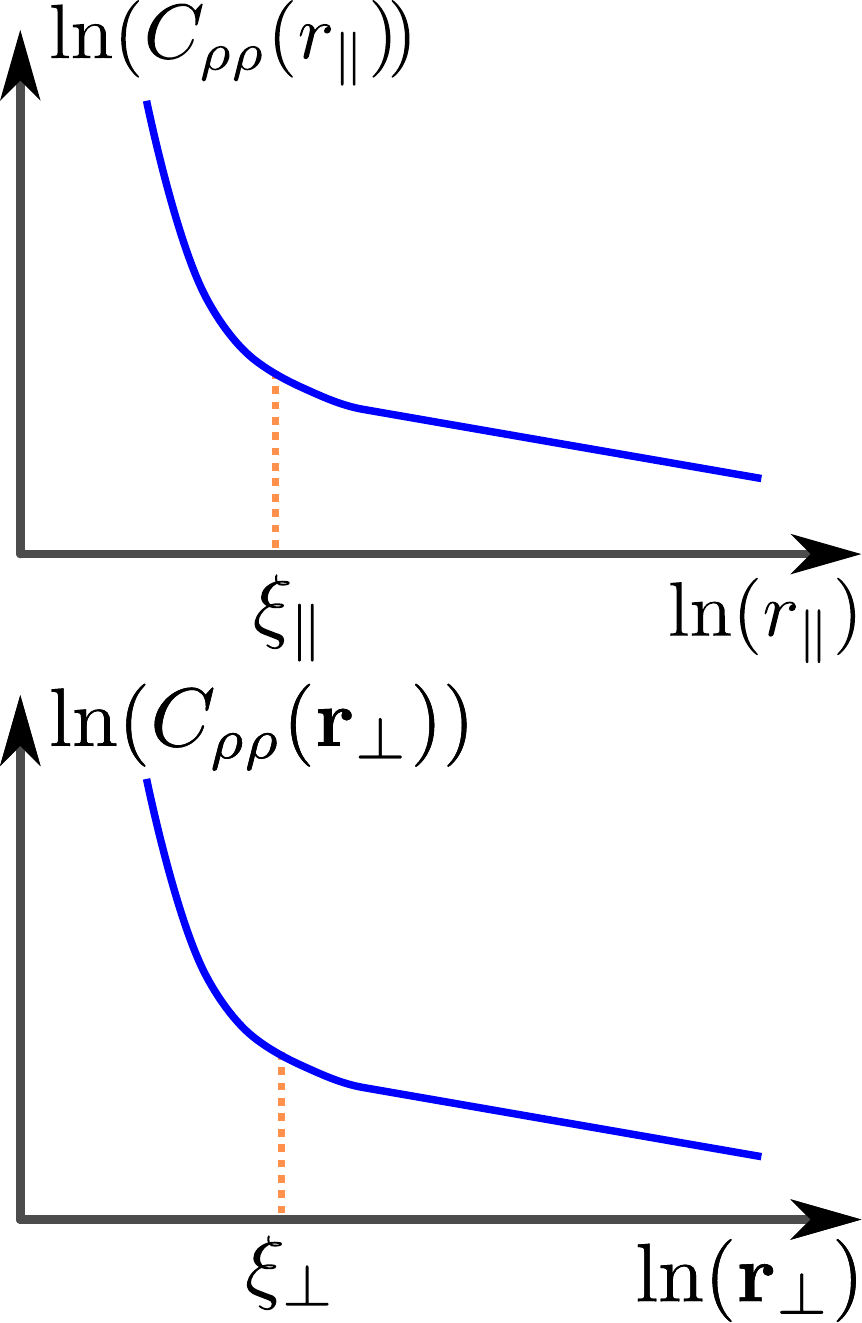}
    \caption{Plots of the realspace scaling of the natural log of the correlation functions $\ln{C_{\rho\rho}(r_\parallel})$ and $\ln{C_{\rho\rho}(r_\perp})$.}
    \label{CorrelationScalingFig}
\end{figure}

In contrast, in phase separating flocks, the correlation lengths $\xi _{\perp,\parallel}$ are the length scales at which the density-density correlation function $C_{\rho\rho}(\br)$ crosses over from one power law decay to another. That is

\begin{align}
    C_{\rho\rho}(\br_\perp, r_\parallel=0) \propto \begin{cases}
        r_\perp^{2\chi_\perp^c} &,~ r_\perp<<\xi_\perp \vspace{0.2cm}\\
        r_\perp^{2\chi_\perp} &,~ r_\perp>>\xi_\perp \label{rho corr perp}
    \end{cases}
\end{align}

for points separated in the direction perpendicular to the direction "($\parallel$)" of 
 mean flock motion, and
\begin{align}
    C_{\rho\rho}(\br_\perp={\bf 0}, r_\parallel) \propto \begin{cases}
        r_\parallel^{2\chi_\parallel^c} &,~ r_\parallel<<\xi_\parallel \vspace{0.2cm}\\
        r_\parallel^{2\chi_\parallel} &,~ r_\parallel>>\xi_\parallel \label{rho corr par}
    \end{cases}
\end{align}
for points separated along the direction of 
 mean flock motion. This behavior is illustrated in figure (\ref{CorrelationScalingFig}).
 
Our $\epsilon$-expansion results for the critical ``roughness exponents" $\chi^c_{\perp,\parallel}$ in spatial dimensions $d=5-\epsilon$ are
\begin{align}
    \chi_\perp^c &= -2 + \frac{\epsilon}{2} + \order(\epsilon^2)\\
    \chi_\parallel &= -4+ \epsilon + \order(\epsilon^2)
\end{align}

In $5-\epsilon$ dimensions, the {\it non}-critical roughness exponents are given by

\begin{align}
    \chi_\perp &= \frac{3-d}{2} = -1 + \frac{\epsilon}{2}\label{chi1}\\
    \chi_\parallel &= 3-d = -2 + \epsilon \label{chi2}
\end{align}
 {\it exactly}. However, once the spatial dimension goes below $d=4$ (which is obviously true for all physically relevant cases), the exact results
{ (\ref{chi1}) and (\ref{chi2})} cease to hold. The $\chi$'s that apply then are simply those of the ordered state of a polar active fluid (i.e., a "flock") in that particular dimension of space. These exponents are known only from simulation,\cite{ginpreprint} and are given in three dimensions by 
\beq
\chi_\perp\approx\chi_\parallel \approx-.62 \,.
\label{chi3}
\eeq

The remainder of this paper is organized as follows. In section (\ref{hyd}), we present the hydrodynamic theory of phase separating flocks near the critical point. We solve the 
linearized version of this theory in section (\ref{lin}). In section (\ref{NL}), we analyse the full model, including non-linearities, using the dynamical renormalization group (DRG), and thereby show that this linear theory breaks down for spatial dimensions $d\le5$, in contrast to equilibrium phase separation, for which this breakdown only occurs for $d\le4$. We use the DRG results of section (\ref{NL}) to find the fixed point that controls the new universality class of phase separating flocks, and determine the critical exponents $\beta$, $\nu_{\perp,\parallel}$,  and $\chi^c_{\perp, \parallel}$ to linear order in $\epsilon=5-d$. Section (\ref{con}) summarizes our results.

\vspace{0.2cm}
\section{Hydrodynamics}\label{hyd}
Our  hydrodynamic model of a generic polar active fluid near the critical point for phase separation is the Toner-Tu theory of flocks \cite{TT1,TT2,TT3,TT4,TT5}, with a few small but important modifications.  The most crucial difference is that we consider the case in which the inverse compressibility (defined precisely below) of the flock is tuned through zero to negative values.  

The theory is a  continuum model for two fields: the number density of boids $\rho(\br,t)$, and  the boid velocity field $\bv(\br,t)$.  The equations of motion for these fields are:

\bew
\begin{align}
    \nonumber &\pp_t\bv+ \lambda_1(\bv\cdot\nabla)\bv + \lambda_2(\nabla\cdot\bv)\bv + \lambda_3\nabla(|\bv|^2) = \\
    &\quad \quad U(|\bv|,\rho)\bv - \nabla P_1(|\bv|,\rho) - \bv(\bv\cdot\nabla P_2(|\bv|,\rho)) + D_B\nabla(\nabla\cdot\bv) + D_T\nabla^2\bv + D_2(\bv\cdot\nabla)^2 \bv + \bff \label{EOMVel2-1}\\
    &\pp_t\rho + \nabla\cdot(\bv\rho) = \nabla\cdot{\bf f}_\rho  \label{EOMDen2-2}
\end{align}
\ew

The significance of these terms is as follows:

The terms involving the parameters $\lambda_i$ are analogs of the convective derivative of the coarse grained velocity field from the Navier-Stokes equations. If our system respected Galilean invariance, we would have $\lambda_1 = 1$ and $ \lambda_{2,3} = 0$. However, because our flock is on a frictional substrate, which provides a special reference frame, we have neither Galilean invariance nor momentum conservation.

\begin{figure}
    \centering
    \includegraphics[width=0.95\linewidth]{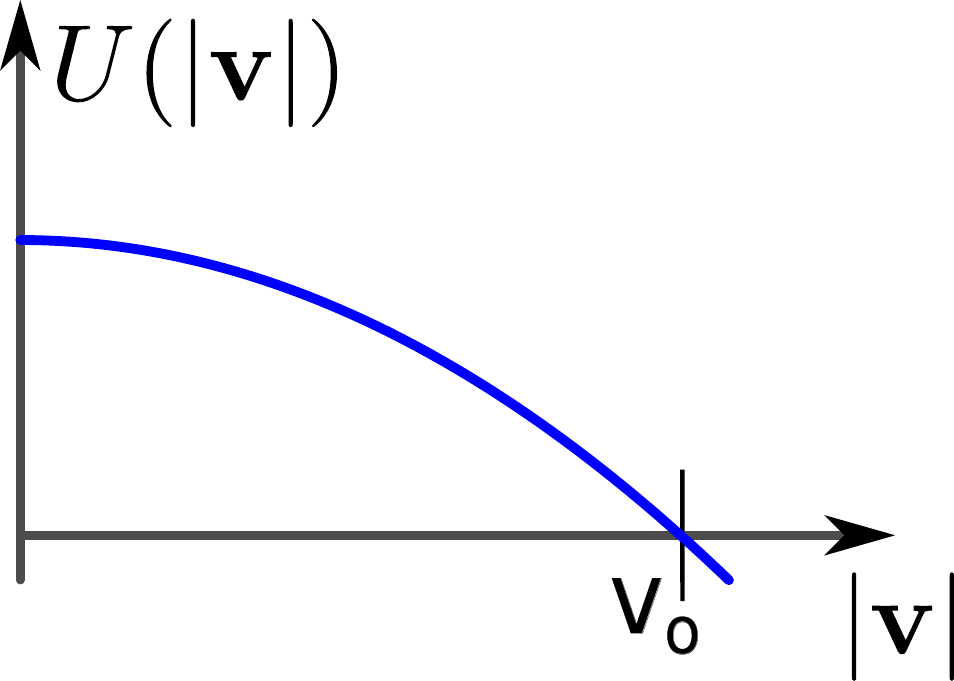}
    \caption{ A plot of $U(|\bv|, \rho)$ for fixed $\rho$. The effect of this term is to maintain a non-zero speed, $v_0$, of the flock. Boids will accelerate if their speed is below $v_0$ and decelerate if their speed is larger than $v_0$.}
    \label{disspationgraph}
\end{figure}

The $U(|\bv|,\rho)$ term, which is similar in form to  a dissipative term, clearly therefore also breaks both Galilean invariance  and momentum conservation. However, because our system is active $U(|\bv|,\rho)$ need not be negative for all $|\bv|$; indeed, if we are to model a system in which the steady state is a {\it moving} flock, we must take it to have the form illustrated in figure  \ref{disspationgraph}. In earlier literature, the special choice $U(|\bv|,\rho)=A - u|\bv|^2$ is often made. This is by no means necessary, however.  

The $P_1$ term is perfectly analogous to  the isotropic pressure in the Navier-Stokes equations. The $P_2$ term is an ``anisotropic pressure", and is allowed because our system breaks rotation invariance locally, which means that there is no reason that the response to density gradients {\it along} the local velocity $\bv$ should be the same as that to gradients perpendicular to $\bv$. Note that this term also breaks Galilean invariance.

The  velocity diffusion constants $D_B$ and  $D_T$ are precise analogs of the bulk and shear viscosities, respectively, in the Navier-Stokes equations. The diffusion constant $D_2$ is an anisotropic viscosity which has no analog in the Navier-Stokes equation, because it violates Galilean invariance. Since we lack Galilean invariance here, it is allowed in our problem.
All of these viscosities have the effect of suppressing fluctuations of the velocity away from spatial uniformity.

The quantities $P_{1,2}(|\bv|,\rho)$, $U(|\bv|,\rho)$, $D_{B,T,2}$, and $\lambda_{1,2,3}$ are in general functions of $|\bv|$ and $ \rho$. They can {\it not} depend of the {\it direction} of $\bv$ due to rotation invariance.

We will expand these parameters in powers of $\delta\rho(\br,t)\equiv\rho(\br,t)-\rho_c$, where $\delta\rho(\br)\equiv \rho(\br,t)-\rho_c$ is the departure of the local number density of flockers $\rho(\br,t)$ at position $\br$ and time $t$ from its "critical" value $\rho_c$, and $\delta |\bv(\br,t)|\equiv |\bv(\br,t)|-v_0$ is likewise the departure of the local speed of the flockers $|\bv(\br,t)|$ at position $\br$ and time $t$ from its mean value $v_0$.

Particularly crucial is the expansion of the isotropic pressure $P_1(\rho, |\bv|)$. This can be written
\begin{align}
P_1 = P_0+\frac{m}{\rho_c}\delta\rho  + u \delta\rho^3  -K\nabla^2\delta\rho\,.
\label{Pexp}
\end{align}

This expansion differs from that considered in reference \cite{rean} in a few details, because our interest here is in the critical point, whereas \cite{rean} dealt with the ordered phase. These differences are:

\noindent1) We do not have a quadratic term (i.e., a $\delta\rho^2$ term) in the isotropic pressure $P_1$, because we are expanding around the critical point at which that term vanishes. Reference \cite{rean} had such a term, because it was studying flocks away from the critical point.   In \cite{uslong}, it was shown that the special density around which the expansion of the pressure lacks this quadratic term is the
critical density of the phase separation.

Thus, our leading order non-linearity in density is $O(\delta\rho^3)$.

\noindent2) We will explicitly consider the limit in which the coefficient $m$ of the linear term in $P_1$  vanishes; that is, the limit $m\to0$, which was not considered in \cite{rean}.

\noindent3) The term $K \nabla_\perp\nabla^2\delta\rho$ was not considered in \cite{rean}, because this term is negligible at long wavelengths in the ordered phase, since it scales in Fourier space like $q_\perp^3\delta\rho$, while the $\nu_x\pp_x\nabla_\perp \delta\rho$ scales like $q_xq_\perp\delta\rho$. Since the important regime of wavevectors in the ordered phase is $q_x\sim q_\perp^\zeta$ with $\zeta<1$, the $\nu_x$ term dominates throughout the dominant regime of wavevector. Near the critical point, however, as we show in this paper, the dominant regime of wavevector has $q_x\sim q_\perp^2$ (to one loop order), and so the $\nu_x$ and $K$ terms are both of order $q_\perp^3\delta\rho$. Hence the $K$ term should be kept. 

{ Maxx: I've just checked, and I think that the net effect of adding this term is to change the coefficient of $q_\perp^4$ in our expression for $a(\bq, \omega)$ (which I've renamed $\mu(\bq, \omega)$) from $D_{L\perp}D_{\rho\perp}$ to $D_{L\perp}D_{\rho\perp}+K$. Please check and see if you agree with that statement. If so, please make the necessary  changes throughout the paper (I have already attempted to do so, and found very few places where they were needed, but that could just be because I'm sloppy!). }

As shown in ref \cite{uslong}, the coefficient $m$ of the linear term in this expression can be driven from positive to negative values by increasing autochemotaxis\cite{uslong}, as well as a variety of other mechanisms. Here we will simply treat $m$ as an experimentally tunable parameter that plays a role closely analogous to that of temperature in the liquid-gas phase diagram.

The noise terms ${\bf f}$ and ${\bf f_\rho}$ are assumed to be Gaussian white noise with the correlations:
\begin{align}
    \langle f_i(\br,t) f_j(\br',t')\rangle = \Delta \delta_{ij} \delta^d(\br - \br')\delta(t-t')\label{velForceNoiseCorr}\\
     \langle f_{\rho i}(\br,t) f_{\rho j}(\br',t')\rangle = \Delta_\rho \delta_{ij} \delta^d(\br - \br')\delta(t-t')\label{denForceNoiseCorr}
\end{align}

 In reference \cite{uslong}, the noises were set to zero. In a real system, they will always be non-zero. This causes fluctuations, which, as we'll show below, invalidate, for all spatial dimensions $d\le5$,  the ``linear" or, equivalently, ``mean-field",  approach used in reference \cite{uslong}.

We begin by expanding these equations of motion about  a homogeneous moving flock state at the critical density. In a homogeneous state, both of our fields are constants, so we have 
\beqn
    \rho(\br,t) &= \rho_c \,,\label{ss1}\\
        \bv &= v_0\hat{x} \,.
    \label{ss3}
\eeqn
Note that the direction of $\bv$ is completely arbitrary, due to the rotation invariance of our model. We will henceforth choose our coordinate system so that the $x$-axis is the direction of the spontaneous velocity.

Inserting these constant ans\"{a}tze (\ref{ss1}) and (\ref{ss3}),  into our equations of motion  (\ref{EOMVel2-1}) and (\ref{EOMDen2-2}), it is clear that
all terms involving spatial or temporal derivatives vanish. It is easy to see that this implies that the density equation (\ref{EOMDen2-2}) is automatically satisfied. 
The velocity equation reduces to 
\beq
 U(v_0, \rho_c)=0 \,.
 \label{vcond}
 \eeq
 This is a scalar algebraic equation for the unknowns $\rho_c$ and $v_0$. We clearly need one more condition. This can be obtained by fixing $\rho_c$ to be the value of $\rho$ at which the expansion of the isotropic pressure $P_1$ takes the form \rf{Pexp}; that is, that there is no term quadratic in $\delta\rho$.

 With these values of $\rho_c$ and $\eta_0$ in hand, we can then in principle use (\ref{vcond}) to determine the steady-state speed $v_0$. We will also assume that the solution of (\ref{vcond}) is unique, which it clearly will be if  $U(v_0, \rho_c, \eta_0)$ looks like figure  \ref{disspationgraph}.

We now expand our equations of motion about this steady state solution. That is, we will write
 \beqn
    \rho(\br,t) &=& \rho_c + \delta\rho(\br,t),\\
    \bv(\br,t) &=& v_0\hat{x} + {\bf  \delta v}(\br,t) = (v_0 + \delta v_x(\br,t))\hat{x} + \bv_\perp(\br,t),\nn\\
\label{field_expand}
\eeqn
and then expand our equations of motion to cubic order in the fluctuation $\delta\rho(\br,t)$ of the density,  and quadratic order in the fluctuation ${\bf  \delta v}(\br,t)$ of  $\bv(\br,t)$. 
The DRG analysis that we will perform in section (\ref{NL}) shows that this order is sufficient to obtain the universal scaling laws of the transition.

The expansion process of linearization begins by expanding all of the $|\bv|$ and  $\rho$  dependent parameters in the equations of motion. One also needs to eliminate  
the ``fast" mode $\delta v_x$. This process is done explicitly in excruciating detail in reference \cite{uslong}. The only differences between our analysis here and that done in reference \cite{uslong} are the differences in the starting model discussed above.
The result is the following closed set of equations of motion 
 for $\delta\rho$ and $\bvp$:

\begin{widetext}
\begin{align}
\partial_{t} \bv_\perp + v_v\partial_x
\bv_\perp + \lambda \left(\bv_\perp \cdot
\nabla_{\perp}\right) \bv_\perp &=-g_1 \delta\rho\partial_x
\bv_\perp-g_2\bv_\perp\partial_x \delta\rho  - \nabla_\perp \bigg(\frac{m}{\rho_c}\delta\rho  + u \delta\rho^3 { -K\nabla^2\delta\rho}\bigg)
+ D_{B}\nabla_\perp(\nabla_\perp\cdot\bv_\perp) \nonumber\\&+ D_T\nabla^2_\perp\bv_\perp  
+ D_x\partial_x^2\bv_\perp+\nu_x\pp_x\nabla_\perp \delta\rho+\nu_t \pp_t\nabla_\perp \delta\rho + {\bf f}_\perp \,,  
\label{TT v Perp Full}\\\nn\\
\partial_t \delta\rho +\rho_c\nabla_\perp\cdot\bv_\perp
+{\gamma}\nabla_\perp\cdot(\bv_\perp \delta\rho)+v_\rho
\partial_x\delta\rho &= D_{\rho x}\partial^2_x \delta\rho+D_{\rho\perp}\nabla_\perp^2\delta \rho
+D_{\rho v} \partial_x
\left(\nabla_\perp \cdot \vec{v}_{\perp}\right)+\phi\partial_t\partial_x \delta\rho
+w_2\partial_x( \delta\rho^2)\nn\\&+{ \rho_c\over2v_0}\partial_x(|
\vec{v}_\perp|^2) + \nabla\cdot{\bf f}_\rho ~, \label{TT delta rho Full}
\end{align}
\end{widetext}
where we've defined $\lambda\equiv\lambda_1( \rho_c, v_0)$.

Strictly speaking, the coefficient $\gamma$ of the $\nabla_\perp\cdot(\bv_\perp \delta\rho)$ term in the equation of motion \ref{TT delta rho Full} for $\delta\rho$ is $\gamma=1$ at this point. We have introduced $\gamma$ because this coefficient will {\it not} remain equal to $1$ upon renormalization, as we will see when we perform the DRG in section \ref{NL}.

\vspace{0.2cm}

\section{Linear Analysis}\label{lin}

As a first step towards understanding the effect of fluctuations, we will solve these equations of motion to {\it linear} order. Our goal is to solve for $\bv$ in terms of the forces $\bff$ and $\bff_\rho$. Once we have these solutions, we can obtain the correlations of the fields $\delta\rho$ and $\bvp$ from the known correlations (\ref{velForceNoiseCorr}) and (\ref{denForceNoiseCorr}) of the forces. 

To linear order, the equations of motion read:

\bew
\beqn
     \partialder{\bv_\perp}{t} =&-& v_v\partial_x\bv_\perp -\frac{m}{\rho_c}\nabla_\perp \delta \rho{ + K \nabla_\perp \nabla^2} \delta \rho
    + D_{BR}\nabla_\perp(\nabla\cdot\bv_\perp) + D_T\nabla^2_\perp\bv_\perp + D_x\partial_x^2\bv_\perp\nn\\&+&\nu_x\pp_x\nabla_\perp\delta\rho+\nu_t \pp_t\nabla_\perp\delta\rho + {\bf f}_\perp\,,  \label{vperpEq2}\\\nn\\
    \partialder{\delta\rho}{t} = &D_{\rho\perp}&\nabla_\perp^2\delta \rho +D_{\rho x}\pp_x^2\delta \rho- v_\rho\partial_x\delta\rho - \rho_c\nabla_\perp\cdot\bv_\perp+D_{\rho v}\pp_x(\nabla_\perp\cdot\bv_\perp)+\phi\pp_t\pp_x\delta\rho+ \nabla\cdot{\bf f}_\rho \,,  \label{rhoEq2} 
\eeqn
\ew
where all coefficients can be written in terms of the original model parameters.

We then proceed by Fourier transforming these equations of motion, using the following convention for Fourier transforms:
\begin{align}
    f(\bq,\omega) = \int_{t}\int_\br e^{-i(\bq\cdot\br- \omega t)}f(\br,t),
\end{align}
where $\int_t \equiv \int dt$ and $\int_\br \equiv d^dr$. The inverse transform is then given by 
\begin{align}
    f(\br,t) = \int_\omega\int_\bq f(\bq,\omega) e^{i(\bq\cdot\br -\omega t)} \,,
\end{align}
where, here and throughout this paper, we will use the shorthand notation $\int_\omega \equiv \frac{d\omega}{2\pi}$ and $\int_\bq \equiv \int \frac{d^d q}{(2\pi)^d}$. 

The Fourier transformed equations of motion are:
\bew
\begin{align}
    \bigg[-i(\omega-v_\rho q_x) +  D_{\rho\perp}q_\perp^2 +D_{\rho x}q_x^2  -\phi\omega q_x \bigg] \delta\rho &= -i\rho_c\bq_\perp\cdot\bv_\perp + i\bq\cdot{\bm f}_\rho\label{rhoEq3} ,\\
    \bigg[-i\omega + iv_vq_x + D_Tq^2_\perp + D_xq^2_x \bigg]\bv_\perp + D_B \bq_\perp(\bq_\perp\cdot \bv_\perp) &= \bigg[-i\bigg(\frac{m}{\rho_c}{ +Kq_\perp^2}\bigg)+\nu_t\omega-\nu_xq_x\bigg]\bq_\perp\delta\rho +  {\bm f}_\perp \label{vperpEq3}.
\end{align}
\ew

We decouple these by projecting (\ref{vperpEq3}) perpendicular to and along $\bq_\perp$. That is, we write 
\begin{align}
\bv_\perp = v_L\hat{q}_\perp + \bv_T \,,
\end{align} 
with the ``transverse"  components $\bv_T$, by definition, perpendicular to $\bqp$. That is \begin{align}
\bv_T\cdot\hat{q}_\perp = 0,
\end{align} 
and the single ``longitudinal" component 
\begin{align} 
v_L = \frac{\bq_\perp\cdot\bv_\perp}{|q_\perp|},
\end{align} 
is the projection of $\bvp$ onto $\bqp$.

We preform a similar decomposition of the noise $\bff_\perp$:
\begin{align}
\bff_\perp = f_L\hat{q}_\perp + \bff_T \,,
\end{align} 
with \begin{align}
\bff_T\cdot\hat{q}_\perp = 0 \,,
\end{align} 
and 
\begin{align} 
\bff_L = \frac{\bq_\perp\cdot\bff_\perp}{|q_\perp|},
\end{align} 

These expressions can also be conveniently rewritten in terms of the ``transverse" and ``longitudinal" projection operators\cite{projperp}:
\begin{align}
    L_{ij}^\perp (\bqp) &\equiv \frac{q_i^\perp q_j^\perp}{q_\perp^2},\label{LogProj}\\
    P_{ij}^\perp(\bqp) &\equiv \delta_{ij}^\perp - L_{ij}^\perp,\label{PerpProj}
\end{align}
where $\delta^\perp_{ij}=1$ if $i=j$ {\it and} $i\ne x\ne j$, and $\delta^\perp_{ij}=0$ otherwise. The tensor $\delta_{ij}^\perp$ projects any vector perpendicular to the direction of mean flock motion $\hx$, while $P_{ij}^\perp(\bqp)$ projects any vector simultaneously perpendicular to {\it both} $\hx$ {\it and} the projection $\bqp$ of $\bq$ perpendicular to $\hx$. The tensor $ L_{ij}^\perp (\bqp)$ simply projects any vector along $\bq$.

In terms of these operators,
\begin{align}
    v_{Ti} &= P_{ij}^\perp v_j^\perp,\\
    v_L\hat{q}_i^\perp &=  L_{ij}^\perp v_j^\perp \\
        f_{Ti} &=  P_{ij}^\perp f_j^\perp,\\
    f_L\hat{q}_i^\perp &=  L_{ij}^\perp f_j^\perp \,.
\end{align}

This split decouples $\delta\rho$  and $v_L$ from the $d-2$  transverse modes $\bv_T$, as can be seen by projecting equation (\ref{rhoEq3}) and (\ref{vperpEq3}) in the transverse and longitudinal directions, which gives:
\bew
\begin{align}
    \bigg[-i(\omega - v_\rho q_x) + \Gamma_\rho(\bq) -\phi\omega q_x \bigg] \delta\rho &= -i\rho_cq_\perp v_L +   i\bq\cdot\bfr\,, 
    \label{rhoEq4} \\
    \bigg[-i(\omega - v_vq_x) + \Gamma_L(\bq)\bigg] v_L &= \bigg[-i\bigg(\frac{m}{\rho_c}{+Kq_\perp^2}\bigg)+\nu_t\omega-\nu_xq_x\bigg]q_\perp \delta \rho + f_L, \label{vLEq1} \\
    \bigg[-i(\omega - v_vq_x) + \Gamma_T(\bq)\bigg] \bv_T &= {\bm f}_T \,. 
    \label{vTEq1}
\end{align}
\ew

Here we've defined 
\beqn
\Gamma_T(\bq)&\equiv& D_T q^2_\perp + D_xq_x^2 \,,
\label{GammaT}
\nn\\
\Gamma_L &\equiv& D_{_{L\perp}} q^2_\perp + D_xq_x^2 \,,
\label{GammaL}
\nn\\
D_{_{L\perp}}  &\equiv& D_T + D_B \,,
\nn\\
\Gamma_\rho =&\equiv& D_{\rho\perp}q_\perp^2+D_{\rho x}q_x^2 \,.
\label{Gammarho}
\eeqn

 Because $\bv_T$ decouples from $v_L$ and $\dr$, we can immediately read off the solution for the field $\bv_T$ in terms of the forces $\bff_T$:
 \beq
 \bv_T(\bq,\omega)=G_{TT}(\bq, \omega)\bff_T
 \label{vtsol}
 \eeq
 where we've defined the ``transverse propagator"
 \begin{align}
    G_T(\bq, \omega) \equiv \frac{1}{-i(\omega-v_vq_x)+\Gamma_T(\bq)} \,.\label{GTransProp}
\end{align}

We can simplify the remaining equations for $\delta\rho$ and $v_L$ considerably by restricting our attention to  the regime of wavevector $\bq$ and frequency $\omega$ that dominates the fluctuations near the critical point; i.e., at small values of the parameter $m$ in the expansion \ref{Pexp} of the isotropic pressure. As we will show below, in this regime, $\omega\sim q_x\sim q_\perp^2$, and $q_\perp$ is very small (specifically, of order $\sqrt{m}$). In this limit, the $D_{\rho x}q_x^2 $ and  $\phi\omega q_x$ terms in \rf{rhoEq4} are clearly both of order $q_\perp^4$, and, hence, negligible, for small $q_\perp$, relative to the $ D_{\rho\perp}q_\perp^2 $ term. We'll therefore drop those $D_{\rho x}$ and $\phi$ terms. 

Likewise, the $D_xq_x^2$ term in \rf{vLEq1} is also of order $q_\perp^4$, and hence negligible relative to the $D_{_{L\perp}} q^2_\perp$ term. We'll therefore drop the  $D_xq_x^2$ term in \rf{vLEq1} as well.

With these simplifications, the longitudinal velocity $v_L$  and density $\delta\rho$ equations of motion  can be rewritten in matrix form:

\begin{align}
    \bm{M}\begin{pmatrix}
        \delta\rho\\ v_L
    \end{pmatrix} = \begin{pmatrix}
        i\bq\cdot\bfr\\f_L
    \end{pmatrix},
\end{align}
where we've defined the matrix 
\bew
\[
\bm{M} (\bq, \omega)= \left(
\begin{array}{cc}
  [-i(\omega-v_\rho q_x) + D_{\rho \perp}q_{\perp}^2] & i \rho_cq_\perp \\[10pt]
{ \bigg[i\bigg(\frac{m}{\rho_c}{+Kq_\perp^2}\bigg)+\nu_xq_x -\nu_t\omega \bigg] q_\perp}&  [-i(\omega-v_vq_x)+D_{_{L\perp}}q_\perp^2] 
\end{array} \,.
\right)\;
\]
\label{Mdef}
\ew

The solution to this linear system of equations is clearly:
\begin{align}
    \begin{pmatrix}
        \delta\rho\\
        v_L
    \end{pmatrix} = \bm{G} \begin{pmatrix}
        i\bq \cdot {\bf f}_\rho\\
        f_L
    \end{pmatrix} \label{linearSystemSol}
\end{align}

\noindent where we've defined the propagator matrix:
\renewcommand\arraystretch{2}
\bew
    \begin{align}
       \bm{G}\equiv \bm{M}^{-1} &= \begin{pmatrix}
            G_{\rho \rho} & G_{\rho L}\\
            G_{L \rho } & G_{L L}
        \end{pmatrix} =\frac{1}{Det[\bm{M}]} \begin{pmatrix}
            [-i(\omega-v_vq_x)+D_{_{L\perp}}q_\perp^2] & - i \rho_cq_\perp \\
   { -\bigg[i\bigg(\frac{m}{\rho_c}{ +Kq_\perp^2}\bigg)+\nu_xq_x -\nu_t\omega \bigg] q_\perp} & [-i(\omega-v_\rho q_x) + D_{\rho \perp}q_{\perp}^2] \,
        \end{pmatrix},
        \label{Minv}
        \end{align}
        \ew
\renewcommand\arraystretch{1}

with the determinant given by 

    \begin{align}
        Det[M] \equiv \mu(\bq, \omega) + i\kappa(\bq, \omega) +mq_\perp^2  \,,
        \end{align}
        where we've defined { I've changed $b$ to $\kappa$ everywhere, because $b$ is already being used for the RG rescaling factor. And having gone greek with one of $a$ and $b$, I've also done so with the other, replacing $a$ with $\mu$ everywhere. If you have a better choice of names for these (one that doesn't use letters already taken, that is), then feel free to change the names to those.)}
\bew
\begin{align}
        \mu(\bq, \omega)&\equiv -\omega^2 + (v_v+v_\rho)\omega q_x + (D_{L\perp}D_{\rho\perp}{ +K{ \rho_c}})q_\perp^4 - v_\rho v_vq_x^2,\label{DeterminateHelper1}\\
        \kappa(\bq, \omega) &\equiv \bigg[(v_\rho D_{L\perp} + v_vD_{\rho \perp}  -\rho_c\nu_x)q_x - (D_{L\perp} + D_{\rho\perp} - \rho_c\nu_t)\omega\bigg] q_\perp^2  \,.\label{DeterminateHelper2}
    \end{align}
\ew

Note that at the critical point $m=0$, $\mu$ and $\kappa$ are both of order $q_\perp^4$ if $\omega\lesssim \sqrt{D_{L\perp}D_{\rho\perp}}q_\perp^2$ and $q_x\lesssim \sqrt{D_{L\perp}D_{\rho\perp}\over v_\rho v_v}q_\perp^2$, and both much greater than order $q_\perp^2$ if either $\omega$ or $q_x$ are much greater than those limits. We will see in a moment that the fluctuations in both velocity and density are  proportional to ${1\over \mu^2(\bq, \omega)+\kappa^2(\bq, \omega)}$. Hence, the regime 
\beq
\omega\lesssim \sqrt{D_{L\perp}D_{\rho\perp}}q_\perp^2 \sep q_x\lesssim \sqrt{D_{L\perp}D_{\rho\perp}\over v_\rho v_v}q_\perp^2
\label{dom regime}
\eeq
dominates the fluctuations. We will use this fact later to help us assess the relative importance of various terms in our model, and thereby eliminate many of them.

Note also that $ \mu(\bq, \omega)$ and $ \kappa(\bq, \omega)$ are both independent of $m$; we will make much use of this fact in the RG analysis of section \ref{NL}.

Using { ({ \ref{linearSystemSol}}) and ({ \ref{Minv}})}, we can summarize our solutions for the velocity and density fields in terms of the noises:
\begin{align}
    v_L(\bq, \omega) &= G_{LL}(\bq,\omega) f_L(\bq,\omega) + G_{L\rho}(\bq, \omega)i\bq\cdot\bfr(\bq, \omega) \,,\label{vLSolution}\\
    v_{Ti}(\bq, \omega) &= G_T(\bq, \omega)P_{ij}^\perp(\hat{q}_\perp)f_j(\bq, \omega)\label{vTSolution}
\end{align}
and, most importantly, the solution for $\delta\rho(\bq, \omega)$:

\begin{align}
    \delta\rho(\bq,\omega) &= G_{\rho L}(\bq,\omega)f_L(\bq,\omega) + G_{\rho\rho}(\bq,\omega)i\bq\cdot{\bf f}_L(\bq,\omega)\\
    &= G_{\rho j}(\bq,\omega)f_j(\bq,\omega)+G_{\rho\rho}(\bq,\omega)i\bq\cdot{\bf f}_L
(\bq,\omega)
\end{align}
where we've defined
\begin{align}
    G_{\rho j}(\bq, \omega) = \frac{q^\perp_j}{q_\perp}G_{\rho L}(\bq, \omega)
 = -\frac{i\rho_cq^\perp_j}{Det[M]} \label{GrhoV}
\end{align}

Combining our solutions { \ref{vLSolution} and \ref{vTSolution}} for $v_L$ and $\bv_T$, and using the projection operators { \ref{LogProj} and \ref{PerpProj}} introduced earlier, we can obtain the solution for the velocity field:
\bew
\begin{align}
    v_i = [G_{LL}(\bq,\omega)L_{ij}^\perp + G_{TT}(\bq,\omega)P_{ij}^\perp]f_i + iG_{L\rho}(\bq,\omega)L_{ij}^\perp q_\perp f_{\rho j}(\bq,\omega)\equiv G_{ij}(\bq, \omega)f_j + G_{ij}^\rho(\bq,\omega)f_j^\rho
\end{align}
\ew
where we've defined
 \vspace{0.2cm}
\begin{align}
    G_{ij}(\bq,\omega)&=G_{LL}(\bq,\omega)L_{ij}^\perp + G_{TT}(\bq,\omega)P_{ij}^\perp
    \end{align}
    and
    \begin{align}
    G_{ij}^{\rho}&=iG_{L\rho}\frac{q_i^\perp q_j^\perp}{q_\perp}\nn\\
    &={ \left[{{m\over\rc}+{ Kq_\perp^2}+i(\nu_t\omega-\nu_xq_x)\over \mu(\bq, \omega)
    +i\kappa(\bq, \omega)+mq_\perp^2}\right]q_i^\perp q_j^\perp} \,.
\end{align}

 \vspace{0.2cm}
We can now autocorrelate our fields with themselves and cross-correlate them with each other. Doing so for the velocity fields gives:

\bew 
\begin{align}
\langle v_i(\bq,\omega)v_j(-\bq,-\omega)\rangle =~&\bigg[G_{LL}(\bq,\omega)G_{LL}(-\bq,-\omega)L_{ik}^\perp L_{jl}^\perp\ \nn\\
&+G_{T}(\bq,\omega)G_{T}(-\bq,-\omega)P_{ik}^\perp P_{jl}^\perp\bigg]\langle f_k(\bq,\omega)f_l(-\bq,-\omega)\rangle \nn\\
&+ G_{L\rho}(\bq,\omega)G_{L\rho}(-\bq,-\omega)L_{ik}^\perp L_{jl}^\perp{q_\perp^2}\langle f^\rho_k(\bq,\omega)f^\rho_l(-\bq,-\omega)\rangle
\end{align}
\ew
Using the noise correlations ({ \ref{velForceNoiseCorr}}) and ({ \ref{denForceNoiseCorr}}) in this expression, along with the identities
\begin{align}
    P_{ik}^\perp P_{jk}^\perp = P_{ij}^\perp \sep L_{ik}^\perp L_{jk}^\perp = L_{ij}^\perp
\end{align}
 obeyed by the projection operators, and using our earlier results { (\ref{GTransProp}) and (\ref{Minv})} for the propagators $G_{LL}(-\bq,-\omega)$, $G_{L\rho}(\bq,\omega)$, and $G_{T}(\bq,\omega)$, we obtain

\bew
\begin{align}
    C_{ij}(\bq, \omega) &= \langle v^\perp_i(\bq, \omega)v^\perp_j(\bq',\omega')\rangle = C_{L}(\bq,\omega) L_{ij}^\perp + C_{T}(\bq, \omega)P_{ij}^\perp 
    \label{Cij}
    \end{align}
    where we've defined
       \begin{align}
    C_{T}(\bq, \omega) &\equiv \frac{\Delta}{[(\omega - v_vq_x)^2 + \Gamma_T(\bq)^2]}\label{CT}\\
    C_{L}(\bq, \omega) &\equiv\left(\frac{\omega^2 -2v_\rho \omega q_x +v_\rho^2q_x^2+D_{\rho\perp}^2q_\perp^4}{(\mu(\bq, \omega)+mq_\perp^2)^2 + \kappa^2(\bq, \omega)}\right)\Delta + \left(\frac{\nu_x^2q_x^2q_\perp^2 -2\nu_x\nu_t\omega q_xq_\perp^2 +\nu_t^2\omega^2q_\perp^2}{(\mu(\bq, \omega)+mq_\perp^2)^2 + \kappa^2(\bq, \omega)}\right)\Delta_\rho
    \label{CL}
\end{align}
\ew
Note that, since, in the dominant regime of wavevector and frequency, $\omega\sim q_x\sim q_\perp^2$, the  $\Delta$ term in \rf{CL} is (near the critical point $m=0$) of order 
$q_\perp^4\over(\mu^2(\bq, \omega)+\kappa^2(\bq, \omega))$, while the $\Delta_\rho$ term is of order $q_\perp^6\over(\mu^2(\bq, \omega)+\kappa^2(\bq, \omega))$; that is, smaller by a factor of $q_\perp^2$. Hence, this term is negligible, and we shall henceforth drop it. 

This leaves us with
\bew
\begin{align}
C_{ij}(\bq, \omega) = \langle \bv_i(\bq, \omega)\bv_j(\bq',\omega')\rangle = C_{L}(\bq,\omega) L_{ij}^\perp + C_{T}(\bq,\omega) P_{ij}^\perp \,.
\end{align}
\ew
\bew
\begin{align}
    C_{\rho\rho}(\bq,\omega) &= G_{\rho L}  (\bq,\omega) G_{\rho L}(-\bq,-\omega)\langle|f_L(\bq,\omega)|^2\rangle + G_{\rho\rho}(\bq,\omega)G_{\rho \rho}(-\bq,-\omega)\langle|f_\rho(\bq,\omega)|^2\rangle\\
    &=\left( \frac{\rho_c^2q_\perp^2}{(\mu(\bq, \omega)+mq_\perp^2)^2 + \kappa^2(\bq, \omega)}\right)\Delta+\left(\frac{\omega^2 -2v_v\omega q_x +v_v^2q_x^2+D_{L\perp}^2q_\perp^4}{(\mu(\bq, \omega)+mq_\perp^2)^2 + \kappa^2(\bq, \omega)}\right)\Delta_\rho 
    \label{Crho}
\end{align}
\ew
Once again power counting in the dominant regime of wavevector and frequency $\omega\sim q_x\sim q_\perp^2$, we see that the  $\Delta$ term in \rf{Crho} is of order $q_\perp^2/(\mu^2(\bq, \omega)+\kappa^2(\bq, \omega))$, while the $\Delta_\rho$ term is of order $q_\perp^4/(\mu^2(\bq, \omega)+\kappa^2(\bq, \omega))$; that is, smaller by a factor of $q_\perp^2$. Hence, this term is also negligible, and we shall henceforth drop it as well. This means that the noise $\bff_\rho$ in the density equation has dropped out of the problem. 

Note that all of the correlation functions we've just found are controlled entirely by the model parameters $D_T$, $D_{L\perp}$, $D_{\rho\perp}$, $v_v$, $v_\rho$, $\rho_c$, $\nu_x$, $\nu_t$, and $\Delta$. We will exploit this fact in our Renormalization Group (RG) treatment in the next section, where will assess the importance of the {\it non-linear} terms in the equations of motion by choosing our RG rescaling factors to keep all of the above parameters fixed upon rescaling.

With the above results { \ref{CT}}, { \ref{CL}}, and { \ref{Crho}} for the spatiotemporally Fourier transformed correlation functions in hand, we can now Fourier transform back to real space and time to obtain the position and time dependent correlation functions. We will focus here on the {\it critical} correlation functions; that is, those when the linear coefficient $m$ in the expansion { \ref{Pexp}} for the isotropic pressure $P_1$ vanishes; that is, when $m=0$.

In this limit, the density-density correlation function $C_{\rho\rho}(\br, t)\equiv \langle\delta\rho(\br, t)\delta\rho({\bf 0}, 0)\rangle$ in real space $\br$ and time $t$ is given by:
\bew
\begin{align}
    C_{\rho\rho}(\br,t) &= \int\frac{d^{d-1}q_\perp}{(2\pi)^{d-1}}\int_{-\infty}^{\infty}\frac{dq_x}{2\pi}\int_{-\infty}^{\infty}\frac{d\omega}{2\pi}C_{\rho\rho}(\bq, \omega)e^{i(\bq\cdot\br-\omega t)}\\
    &= \rho_c^2\Delta\int\frac{d^{d-1}q_\perp}{(2\pi)^{d-1}}\int_{-\infty}^{\infty}\frac{dq_x}{2\pi}\int_{-\infty}^{\infty}\frac{d\omega}{2\pi} \frac{e^{i(\bq_\perp\cdot\br_\perp + q_xr_x-\omega t)}q_\perp^2}{\mu^2(\omega,\bq_\perp,q_x) + \kappa^2(\omega,\bq_\perp,q_x)}\label{CRhoRhorealspace1}
\end{align}
\ew
We can now tease out the dependence of this correlation function on $\brp$, $x$, and $t$ by rescaling variables of integration from $\omega$ and $q_x$  to new variables of integration $\Omega$  and $Q_x$  defined by 
\beq
    \omega \equiv D_Tq_\perp^2\Omega \sep q_x \equiv \frac{D_Tq_\perp^2 }{\sqrt{v_vv_\rho}} Q_x  \,,
    \label{rescale omega qx}
\eeq
where we've defined 
\beq
D_T\equiv\sqrt{D_{L\perp}D_{\rho\perp}+K{ \rho_c}} \,.
\label{dtdef}
\eeq
Making these changes of variable in our expressions { \ref{DeterminateHelper1}} and { \ref{DeterminateHelper2}} for 
$\mu(q_x, \bqp, \omega)$ and $\kappa(q_x, \bqp, \omega)$ gives
    \begin{align}
        \mu(q_x, \bqp, \omega) &=D_T^2q_\perp^4 \alpha(\Omega,Q_x) \,,\\
        \kappa(q_x, \bqp, \omega) &= D_T^2q_\perp^4 \sigma(\Omega,Q_x)  \,,
    \end{align}
 where we've defined
    \begin{align}
        \alpha(Q_x,\Omega) &\equiv -\Omega^2 +\varTheta\Omega Q_x -Q_x^2+1 \,,\\
        \sigma(Q_x,\Omega) &\equiv\varXi Q_x - \varUpsilon\Omega  \,,
        \label{asdef}
    \end{align}
with the dimensionless parameters 
\beqn
\varTheta&\equiv&\sqrt{v_v\over v_\rho}+\sqrt{v_\rho\over v_v} \\
\varXi&\equiv&{(v_\rho D_{L\perp} + v_vD_{\rho \perp}  -\rho_c\nu_x)\over D_T\sqrt{v_vv_\rho}}\\
   \varUpsilon&\equiv&{(D_{L\perp} + D_{\rho\perp} - \rho_c\nu_t)\over D_T}
   \label{dlessdefs}
   \eeqn
Note that $\varTheta$ is bounded below by $2$, while $\varXi$ and $\varUpsilon$ can take on any positive value.

With the change of variables \rf{rescale omega qx}, our expression (\ref{CRhoRhorealspace1}) for the real space density-density  correlation function becomes
\bew
\begin{align}
    C_{\rho\rho}(\br,t) &= { \rho_c^2\Delta\over D_T^2}\int\frac{d^{d-1}q_\perp}{(2\pi)^{d-1}}\int_{-\infty}^{\infty}\frac{dQ_x}{2\pi}\int_{-\infty}^{\infty}\frac{d\Omega}{2\pi} \frac{e^{i(\bq_\perp\cdot\br_\perp + q_xaq_\perp^2r_x-\Omega D_Tq_\perp^2t)}}{q_\perp^2\bigg(\alpha^2(Q_x, \Omega) + \sigma^2(Q_x, \Omega)\bigg)} \,.
\end{align}
\ew
where $a\equiv \frac{D_T }{\sqrt{v_vv_\rho}}$ is a non-universal microscopic length.
Now making one further change of variables
\begin{align}
   \bq_\perp \equiv \frac{\bQ_\perp}{|\br_\perp|}    \,,
\end{align}
gives
\beq
C_{\rho\rho}(\brp, r_x, t)= |r_\perp|^{3-d} f_{\rho \rho}\bigg(\frac{r_x}{|r_\perp|^2},\frac{|t|}{|r_\perp|^2}\bigg) 
\label{crhorho real}
    \eeq
    where we've defined the scaling function
    \bew
\beq
   f_{\rho\rho}(u, \tau) = {\rho_c^2\Delta\over D_T^2}\int\frac{d^{d-1}Q_\perp}{(2\pi)^{d-1}}
   \int_{-\infty}^{\infty}\frac{dQ_x}{2\pi}\int_{-\infty}^{\infty}\frac{d\Omega}{2\pi}\left(\frac{e^{i(\bQ_\perp\cdot\hat\br_\perp + Q_xaQ_\perp^2u-\Omega \tau)}Q_\perp^2}
   {Q_\perp^2\bigg(\alpha^2(Q_x, \Omega) + \sigma^2(Q_x,\Omega)\bigg)}\right) \,.
   \label{scaling function}
\eeq
\ew

Note that if define a "roughness exponent" $\chi_\rho$, an "anisotropy exponent" $\zeta$, and a "dynamical scaling exponent" $z$,    via the definition
 \beq
C_{\rho\rho}(\brp, r_x, t)= |r_\perp|^{2\chi_\rho} f_{\rho \rho}\bigg(\frac{r_x}{|r_\perp|^\zeta},\frac{|t|}{|r_\perp|^z}\bigg)  \,,
\label{crhorho real gen scaling}
    \eeq
then equation \rf{crhorho real} implies that the linear theory predicts that $\chi_\rho=\chi_{_{\rho \,{\rm lin}}}$, $\zeta=\zeta_{_{\rm lin}}$, and $z=z_{_{\rm lin}}$, with
\beq
\chi_{_{\rho \,{\rm lin}}}={3-d\over2} \sep \zeta_{_{\rm lin}}=z_{_{\rm lin}}=2 \,.
\label{lin exps}
\eeq
We will show in the next section that, while the scaling {\it form} \rf{crhorho real gen scaling} continues to hold for spatial dimensions $d<5$, the true values of the exponents $\chi_\rho$,  $\zeta$, and  $z$ are all changed by non-linearities in that range of dimension.

 \vspace{0.2cm}
\section{Nonlinear Regime: Rg analysis}\label{NL}

\subsection{Identifying the relevant vertex, and determining the critical dimension}\label{vertex}

Now we wish to go beyond the linear theory presented in the last section, and include the effects of the non-linear terms in  the full equations of motion { (\ref{TT v Perp Full}) and (\ref{TT delta rho Full})}.  
We'll do so  using a dynamical renormalization group (hereafter DRG) analysis.  Readers interested in a more complete and pedagogical  discussion of the DRG are referred to \cite{FNS} for the details of this general approach, including the use of Feynman graphs in it.

First we decompose the Fourier modes $\bvp(\bq, \omega)$ and $\rho(\bq, \omega)$ into  rapidly varying parts $\bvp^>(\bq, \omega)$  and $\rho^>(\bq, \omega)$ and   slowly varying parts $\bvp^<(\bq, \omega)$ and $\rho^<(\bq, \omega)$  in the equations of motion { (\ref{TT v Perp Full}) and (\ref{TT delta rho Full})}. The rapidly varying part is supported in the momentum shell $-\infty <k_x<\infty$, $\Lambda b^{-1}<k_{\perp}<\Lambda$, where $\dd\ell$ is an infinitesimal
and $\Lambda$ is the ultraviolet cutoff. The slowly varying part is supported in $-\infty <k_x<\infty$, $0<k_{\perp}<b^{-1}\Lambda$, where $b$ is an arbitrary rescaling factor that we will ultimately take to be $b=1+d\ell$, with $d\ell$ differential. We separate the noise $\bff$ in exactly the same way.

The  DRG procedure then consists of two steps. In step 1, we eliminate $\bvp^>(\bq, \omega)$ and $\rho^>(\bq, \omega)$ from { (\ref{TT v Perp Full}) and (\ref{TT delta rho Full})}.
We do this by solving iteratively for $\bvp^>(\bq, \omega)$ and $\rho^>(\bq, \omega)$. This solution is a perturbative expansion in $\bvp^<(\bq, \omega)$ and $\rho^>(\bq, \omega)$, as well as the fast components $\bff^>$ of the noise. As usual, the perturbation theory can be represented by Feynman graphs. We substitute these solutions into { (\ref{TT v Perp Full}) and (\ref{TT delta rho Full})} and average over  the short wavelength  components $\bff^>(\bq, \omega)$ of the noise $\bff$, which gives a closed EOM  for $\bvp^<(\bq, \omega)$. 

In Step 2, 
we rescaling the {\it real space} fields $\bvp^<(\br, t)$ and $\rho^<(\br, t)$, time $t$, and coordinates $\brp$ and $r_x$ as follows:
 \beqn
 \bv_\perp &=& b^\chi \bv_\perp' \sep \delta\rho = b^{\chi_\rho}\delta\rho' \sep t=b^zt' \sep \br_\perp=b\br_\perp' \nn\\ r_x &=& b^\zeta r_x' \,.
 \label{rescaling}
 \eeqn
 The rescaling of $\brp$ 
 restores the ultraviolet cutoff back to $\Lambda$. The velocity and density rescaling exponents $\chi$ and $\chi_\rho$, the ``dynamical" exponent $z$, and the ``anistropy" exponent $\zeta$, are all, at this point, completely arbitrary. We will show below that, as usual in RG calculations, there is a particular choice of all of these expoenents that makes it particularly simple to determine which non-linearities are important.  
 
 After these two RG steps, we reorganize the resultant EOMs so that they have the same form as { (\ref{TT v Perp Full}) and (\ref{TT delta rho Full})}, but with various coefficients renormalized.
 
 We focus first on the coefficients that control the size of the fluctuations in the linear theory. Upon performing
the two steps on our equation of motion { (\ref{TT v Perp Full}) and (\ref{TT delta rho Full})}, and the aforementioned reorganization (which amounts to multiplying the EOM by a power of $b$ chosen to restore the coefficient of $\pp_t \bvp$ and $\pp_t\delta\rho$ in { (\ref{TT v Perp Full}) and (\ref{TT delta rho Full})} to unity), we find: 

\begin{align}
    v_v' &= b^{z-\zeta} (v_v + {\rm graphs}) \,,\nn\\
    v_\rho' &= b^{z-\zeta} (v_\rho + {\rm graphs}) \,,\nn\\
    D_{L\perp}' &= b^{z-2} (D_{L\perp} + {\rm graphs}) \,,\nn\\
    D_{\rho\perp}' &= b^{z-2} (D_{\rho\perp} + {\rm graphs}) \,,\nn\\
        D_T' &= b^{z-2} (D_T+ {\rm graphs}) \,,\nn\\
    \nu_t' &= b^{\chi_\rho - \chi - 1} (\nu_t + {\rm graphs}) \,,\nn\\
    \nu_x' &= b^{z-\zeta-1-\chi +\chi_\rho} (\nu_x + {\rm graphs}) \,,\nn\\
    \rho_c'&=b^{z+\chi-\chi_\rho-1}\rho_c \nn\\
    \Delta' &= b^{z-2\chi -\zeta -d + 1}(\Delta + {\rm graphs}) \,,\nn\\
        \Delta_\rho' &= b^{-2}(\Delta_\rho + graphs) \,,\nn\\
    D_x'&=b^{z-2\zeta}(D_x+ {\rm graphs}) \,,
    \label{lin rescale}
\end{align}
where ``graphs" denote the ``graphical" corrections coming from the first, perturbative step of the RG.

Now we'll make the convenient choice of the rescaling exponents $\chi$, $\chi_\rho$,  $z$, and $\zeta$ mentioned above. Our choice will be to choose them so that we  keep the parameters listed in 
\rf{lin rescale} fixed. Doing so means we keep the size of the fluctuations fixed, since, as we showed in the previous section, that size, at least in the linear approximation, is controlled entirely by the set of parameters listed in \rf{lin rescale}.

This means that we can determine whether any of the non-linear terms in equations { (\ref{TT v Perp Full}) and (\ref{TT delta rho Full})} become more important as we renormalize simply by asking whether their coefficients grow or shrink upon renormalization.

We will simplify the argument further by assuming that the bare values of those non-linear coefficients are so small that the "graphs" in equation \rm{lin rescale} are negligible. Once we neglect those corrections, keeping the parameters in \rm{lin rescale} fixed can obviously be achieved simply by choosing the rescaling exponents $\chi$, $\chi_\rho$,  $z$, and $\zeta$ so as to make the exponents of the rescaling factor $b$ in  \rm{lin rescale} vanish. Doing so for $v_v$ and $v_\rho$ clearly leads to $z=\zeta$.
Keeping $D_{L\perp}$ and $D_{\rho\perp}$ fixed requires $z=2$. 
Keeping $\nu_t$ fixed implies $\chi_\rho=\chi+1$, while keeping $\Delta$ fixed requires 
$z-2\chi -\zeta -d + 1=0$. To summarize, keeping these six parameters fixed requires the simultaneous conditions
\beq
z=\zeta \sep z=2 \sep \chi_\rho=\chi+1 \sep z-2\chi -\zeta -d + 1=0 \,.
\label{exp cond}
\eeq
These four simultaneous linear equations for the four unknown exponents can be easily solved, with the result:
\begin{align}
    z = \zeta = 2 \,,\quad
    \chi = \frac{1-d}{2} \,,\quad
    \chi_\rho = \frac{3 - d}{2} \,.
    \label{lin exps}
\end{align}
It is straightforward to check that this choice of exponents will keep all of the other parameters listed in \rf{lin rescale} fixed, {\it except} for $D_x$ and $\Delta_\rho$, whose recursion relations, with the choice  \rf{lin exps} become
\beqn
    \Delta_\rho' &=& b^{-2}(\Delta_\rho + {\rm graphs}) \,, \nn\\
     D_x'&=&b^{-2}(D_x+ {\rm graphs}) \,.
     \label{dxrescale}
     \eeqn
     
The recursion relation for  $\Delta_\rho$ shows that it is irrelevant, as we'd already concluded in our linear analysis of section \ref{lin}.
     
The fact that $D_x$ will flow to zero upon renormalization is more problematic. Since the fluctuations of the transverse velocity $\bv_T$ diverge as $D_x\to0$,  $D_x$ is a "dangerously irrelevant" variable in the renormalization group sense. That is, it is not adequate to simply {\it set} it to zero, rather, we must  keep track of {\it how fast} it scales to zero. More precisely, the effect of the dangerous irrelevance of $D_x$ is only to raise the critical dimension of the  velocity-dependent non-linearities
$\lambda$, $g_1$, $g_2$, $\gamma$, and $\frac{\rho_c}{2v_0}$ to $d=4$; hence, they all remain {\it irrelevant} near $d=5$, as does the $w_2$ non-linearity. Therefore, the only relevant non-linearity near $d=5$ is $u\delta\rho^3$. This tremendously simplifies our analysis.

The  nonlinear terms in the equations of motion { (\ref{TT v Perp Full}) and (\ref{TT delta rho Full})} scale as:
\bew
\begin{align}
    u' &= b^{3\chi_\rho+z-1}(u + {\rm graphs})=b^{5-d}(u + {\rm graphs}),\label{uEoM}\\
    \lambda' &= b^{\chi+z-1}(\lambda + {\rm graphs})= b^{(3-d){ /2}}(\lambda + {\rm graphs}),\\
    g_1' &=b^{\chi_\rho+z-\zeta}(g_1 + {\rm graphs})= b^{(3-d){ /2}}(g_1 + {\rm graphs}),\\
    g_2' &=b^{\chi_\rho+z-\zeta}(g_2 + {\rm graphs})= b^{(3-d){ /2}}(g_2 + {\rm graphs}),\\
    w_2' &= b^{\chi_\rho+z-\zeta}(w_2 + {\rm graphs})= b^{(3-d){ /2}}(w_2 + {\rm graphs}),\\
    \bigg( \frac{\rho_c}{2v_0}\bigg)' &= b^{2\chi-\chi_\rho+z-\zeta}\bigg(\frac{\rho_c}{2v_0} + {\rm graphs}\bigg)= b^{-(1+d){ /2}}\bigg(\frac{\rho_c}{2v_0} + {\rm graphs}\bigg),\\
    \gamma' &= b^{\chi+z-1}(\gamma + {\rm graphs})= b^{(3-d){ /2}}(\gamma + {\rm graphs}),
\end{align} 
\ew
where the second equality in each case comes from using the exponents { \ref{lin exps}} in the first equality in each case.

We see from these relations that, as lower spatial dimension $d$ from very high values, all of the non-linearities are irrelevant, until we reach the ``upper critical dimension" $d_c=5$. At and near this point - that is, for $d=5-\epsilon$ with $\epsilon\ll1$, $u$ - {\it and only }$u$, is relevant. All other non-linearities are {\it irrelevant}, and can hence be ignored in a $5-\epsilon$-expansion, to which we will  turn in the next section.

We'll see later that we'll also need the scaling behavior of the ``mass" $m$:
\begin{align}
       \bigg(\frac{m}{\rho_c}\bigg)' &=b^{\chi_\rho-\chi+z-1}\bigg(\frac{m}{\rho_c} + graphs\bigg)= b^{2}\bigg(\frac{m}{\rho_c} + graphs\bigg)
      \,,\nn
       \label{mEoM}\\
\end{align}
 where in the second equality we have again used the values { \ref{lin exps}} of the exponents.

 \vspace{2cm}

\subsection{5-$\epsilon$-expansion}\label{vertex}

We demonstrated in the last subsection\rf{vertex} that the upper critical dimension for our problem is $d_c=5$, and that only the $u$ vertex is relevant near five dimensions.
To proceed further, we need to actually evaluate the graphical corrections. In this section, we do a full RG treatment accurate to linear order in $\epsilon$.

\begin{figure}
	\begin{center}
 		\includegraphics[scale=.6]{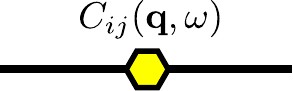}
	\end{center}
	\caption{Rules of graphical representation of propagators, fields, and correlation 
	functions: (a) $=G_{ij}(\tilde{\bk}), G_{\rho j}, G_{\rho\rho}$;	
	(b) $=v^{\perp}_i(\tilde{\bk})$; (c) $=C_{ij}(\tilde{\bk}), C_{\rho j}, C_{\rho\rho}$. }
		\label{FR}
\end{figure}

The basic rules for the graphical representation are illustrated in Fig.~{\ref{FR}}.

\begin{figure}
    \centering
   \includegraphics[width=\linewidth]{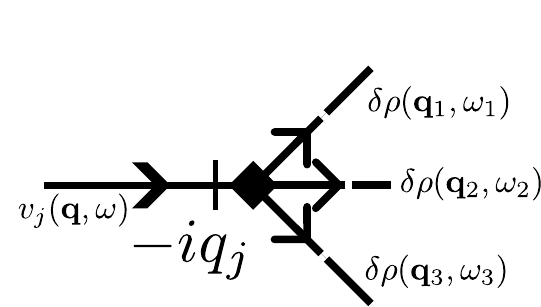}
    \caption{ Graphical representation of the $u$ vertex. The incoming leg on the left labelled $v_i$ tells us that this represents a term in the $v_i$ equation of motion. The tick mark on that leg represents the $q_i$ in the $uq_i\delta\rho^3$ term, and each of the three outgoing legs represents one of those $\delta\rho$'s. See\cite{FNS} for a more detailed description of the Feynman graph representation of equations of motion, and their use in perturbation theory and the DRG.}
    \label{u vert}
    \end{figure}

The $u$ vertex can be represented graphically by the Feynman graph shown in figure \rf{u vert}. The only one loop graphs that can be made from this vertex are the two shown in figures \rf{uGraph} and \rf{uSquaredGraph}. Since the first graph generates a term in the equation of motion for $\bvp$ that is proportional to $\pp_i\delta\rho$, it represents a renormalization of the mass $m$, since $m$ is the coefficient of precisely such a term in that EOM. Likewise, the graph in figure  \rf{uSquaredGraph} is clearly a renormalization of $u$ itself.

\begin{figure}
    \centering
    \includegraphics[width=\linewidth]{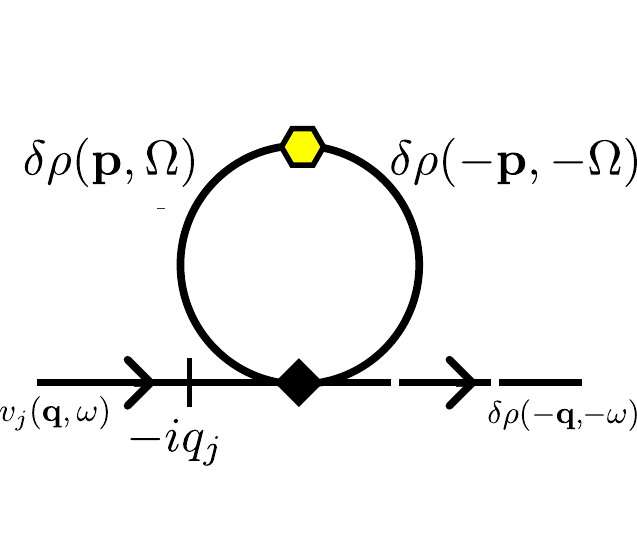}
    \caption{Graphical correction to $m$ from one $u$ vertices. This graph can be made a total of 3 different ways.}
    \label{uGraph}
\end{figure}

\begin{figure}
    \centering
    \includegraphics[width=\linewidth]{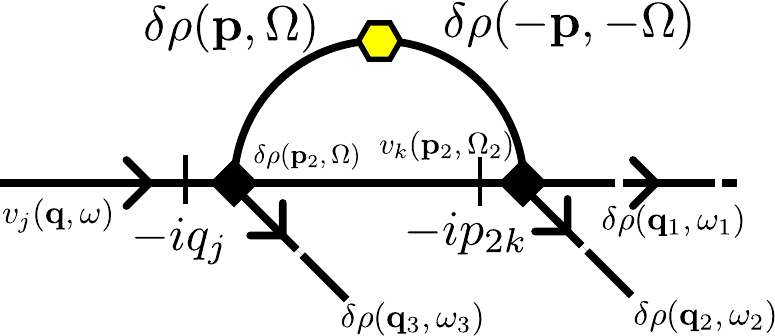}
    \caption{Graphical correction to $u$ from two $u$ verticies. This graph can be made a total of 18 different ways.}
    \label{uSquaredGraph}
\end{figure}

 None of the parameters other than $m$ and $u$ receive any graphical corrections at one loop order.

We'll now evaluate these one loop graphical corrections, starting with figure \rf{uGraph}.  This generates an extra term in the equation of motion for $v_i$ given by 
\begin{widetext}
   
   \beq
   \delta\bigg[ \pp_tv_i\bigg] =  {\rm FT}\bigg\{-3uiq_i\delta\rho(\bq,\omega) \int_{\bp,\Omega} ^>C_{\rho\rho}(\bp, \Omega) \bigg \}\,,
   \label{mcor1}
   \eeq
   \ew
   where ${\rm FT}\{x(\bq,\omega)\}$ denotes a Fourier transform of $x(\bq,\omega)$ back to real space and time; i.e.,  ${\rm FT}\{x(\bq,\omega)\}=x(\br,t)$, and the "$>$" superscript implies that the integral over $\bp$ is over the  the region $b^{-1}\Lambda<|\bpp|<\Lambda$, $-\infty<p_x<\infty$ , which is the region over which we average out the degrees of freedom in each step of our RG.
   
   The form of this generated term is exactly the same as that of the ${m\over\rc}$ term already present in the equation of motion { (\ref{TT v Perp Full})}. Hence, this can be interpreted 
   as a graphical correction $\delta\bigg[ \frac{m}{\rho_c}\bigg]$ to $\bigg[ \frac{m}{\rho_c}\bigg]$ given by
   
\bew
\begin{align}
       \delta\bigg[ \frac{m}{\rho_c}\bigg]= {  \int_{\bp,\Omega} ^>C_{\rho\rho}(\bp, \Omega)=}3u\rho_c^2\Delta &\int_{\bp,\Omega} ^>\frac{p_\perp^2}{(\mu(\bp, \Omega)+mp_\perp^2)^2 + \kappa^2(\bp, \Omega) } \,. \label{BCorrection Integral Full}
    \end{align}
    \ew
  Since only the ratio ${m\over\rc}$ appears in the equation of motion, it is completely arbitrary how much of this renormalization we attribute to a renormalization of $m$ and $\rc$ individually. We will henceforth make the entirely arbitrary choice of attributing all of it to renormalization of $m$; that is, we'll treat $\rc$ as a constant.

 Because we are interested in the behavior around the critical point, which occurs at small $m$, we will expand (\ref{BCorrection Integral Full}) to linear order in $m$.(It is straightforward to show that higher order terms in $m$ only change the critical exponents we find here at $O(\epsilon^2)$, which is beyond the order to which we are working.)   This expansion in $m$ gives
\bew
\begin{align}
    \delta\bigg[ \frac{m}{\rho_c}\bigg] = 3 u \rho_c^2 \Delta \int_{\bp, \Omega}^> \bigg[\frac{p_\perp^2}{\mu^2(\bp, \Omega) +\kappa^2(\bp, \Omega)} -\frac{2\mu(\bp, \Omega)p_ \perp^4 m}{\big(\mu^2(\bp, \Omega) + \kappa^2(\bp,\Omega)\big)^2} + \order(m^2)\bigg] \label{Expanded B integral integrand}
\end{align}
\end{widetext}
Making the same change of variables (\ref{rescale omega qx}) that we made earlier when evaluating the space-time density correlation function  - that is, replacing $\Omega$ and $p_x$ with new variables of integration $\varpi$ and $P_x$ defined via:
\beq
    \Omega \equiv D_Tp_\perp^2\varpi \sep p_x \equiv \frac{D_Tq_\perp^2 }{\sqrt{v_vv_\rho}} P_x  \,,
    \label{rescale Omega px}
\eeq
we can rewrite this as
\bew
\begin{align}
    \delta\bigg[ \frac{m}{\rho_c}\bigg] = {3 u \rho_c^2\Delta\over \sqrt{v_vv_\rho}} \bigg({1\over D_T^2} \int_{\bpp}^>{1\over p_\perp^2}I_1(\varTheta, \varXi, \varUpsilon)-{2m\over D_T^4}\int_{\bpp}^>{1\over p_\perp^4}I_2(\varTheta, \varXi, \varUpsilon)+ \order(m^2)\bigg) \,,
    \label{Expanded B integral integrand}
\end{align}
\end{widetext}
where we've defined the two dimensionless, $O(1)$ integrals
\bew
\beq
I_1(\varTheta, \varXi, \varUpsilon)\equiv\int_{-\infty}^\infty {dP_x\over2\pi} \int_{-\infty}^\infty {d\varpi\over2\pi} \bigg[\frac{1}{\alpha^2(P_x, \varpi) +\sigma^2(P_x, \varpi)} \bigg]\sep
I_2(\varTheta, \varXi, \varUpsilon)\equiv\int_{-\infty}^\infty {dP_x\over2\pi} \int_{-\infty}^\infty {d\varpi\over2\pi} 
\frac{\alpha(P_x,\varpi)}{\big(\alpha^2(P_x, \varpi) +\sigma^2(P_x, \varpi)\big)^2} 
\,.
\label{I12def}
\eeq
\ew
{\it Very} fortunately, it proves to be unnecessary to evaluate these integrals in order to calculate the critical exponents to $O(\epsilon)$. It is sufficient for our purposes to know that $\dio$ and $\dit$ are finite and positive for {\it some} range of the three dimensionless parameters $\varTheta$, $\varXi$, and $\varUpsilon$. One can show that both $\dio$ and 
$\dit$ are finite provided that  $\alp$ and $\sigp$ do not vanish at the same point in the $(P_x, \varpi)$ plane. It is straightforward to show that this leads to the condition
\beq
\varTheta\varXi\varUpsilon>\varXi^2+\varUpsilon^2 \,,
\label{stab cond}
\eeq
which is easily satisfied (indeed, since $\varTheta$ is always greater than $2$, this condition is {\it always} satisfied when $\varXi=\varUpsilon$). The integral $\dio$ is clearly positive definite, since the integrand is. We have established that there is some range of $\varTheta$, $\varXi$, and $\varUpsilon$ in which $\dit$ is positive by brute force numerical integration for a few specific sets of values of the dimensionless parameters $\varTheta$, $\varXi$, and $\varUpsilon$.

Note also that, since we will be choosing our arbitrary RG rescaling exponents to keep the parameters iof the linear theory fixed, and since $\varTheta$, $\varXi$, and $\varUpsilon$ depend {\it only} on the parameters of the linear theory, $\varTheta$, $\varXi$, and $\varUpsilon$ will themselves be constant under the RG. Hence, so will $\dio$ and $\dit$.

The integrals over $\bpp$ in \rf{Expanded B integral integrand}, on the other hand, can be readily evaluated, especially in the limit 
\beq
b=1+d\ell \,,
\label{dell}
\eeq
with $d\ell$ differential. Recalling that the superscript ``$>$" on the integral over $\bpp$ implies that $\bpp$ lies in the thin shell $b^{-1}\Lambda<|\bpp|<\Lambda$, and taking the limit of $d\ell\ll1$, it is easy to see that
\beq
\int_{\bpp}^>{1\over p_\perp^2}=K_{d-1}\Lambda^{d-3}d\ell  \sep \int_{\bpp}^>{1\over p_\perp^4}=K_{d-1}\Lambda^{d-5}d\ell 
\label{pperp ints eval}
\eeq
where we've defined
\beq
K_{d-1}\equiv\frac{S_{d-1}}{(2\pi)^{d-1}} \,,
\label {kddef}
\eeq
with $S_{d-1}$ the surface area of a unit $d-1$-dimensional ball.

 Summarizing these results, the graphical corrections $\delta m$ to $m$ at one loop order are given by:
\bew
\begin{align}
    \delta m= {3 K_{d-1}u \rho_c^3\Delta\over \sqrt{v_vv_\rho}} \bigg({\Lambda^{d-3}I_1(\varTheta, \varXi, \varUpsilon)\over D_T^2} -{2m\Lambda^{d-5}I_2(\varTheta, \varXi, \varUpsilon)\over D_T^4}+ \order(m^2)\bigg)d\ell \,,
    \label{m graph}
\end{align}
\end{widetext}
with $\dio$ and $\dit$ given by \rf{I12def}. Keep in mind that both of these integrals are constant under the RG.

Using this result for the term ``graphs" in equation \rf{mEoM} gives us the complete RG recursion relation for $m$:
\bew
\begin{align}
     m' &=b^{\chi_\rho-\chi+z-1}(m+ \delta m)= b^{\chi_\rho-\chi+z-1}\bigg[m+{3 K_{d-1}u \rho_c^3\Delta\over \sqrt{v_vv_\rho}} \bigg({\Lambda^{d-3}I_1(\varTheta, \varXi, \varUpsilon)\over D_T^2} -{2m\Lambda^{d-5}I_2(\varTheta, \varXi, \varUpsilon)\over D_T^4}+ \order(m^2)\bigg)d\ell\bigg]
      \,.\nn
       \label{mrr1}\\
\end{align}
\ew
The RG now proceeds by iterating this process. The result can be summarized by a differential recursion relation in the following (by now very standard) manner:
as already mentioned, we choose $b=1+d\ell$ with $d\ell$ differential. Instead of keeping track of the number $n$ of iterations of the renormalization group, we introduce a ``renormalization group time'' $\ell$ defined as $\ell\equiv nd\ell$. Then applying the recursion relation { \ref{mrr1}} after a renormalization group time $\ell$, we must evaluate all of the parameters on the right hand side at RG time $\ell$; the result $m'=m(\ell+d\ell)$. Hence, after expanding the right hand side of (\ref{mrr1}) to linear order in $d\ell$, we can rewrite (\ref{mrr1}) as
\bew
\begin{align}
     m(\ell+d\ell) &=m(\ell)+\bigg[(\chi_\rho-\chi+z-1)m(\ell)+ {3 K_{d-1}u(\ell) \rho_c^3\Delta\over \sqrt{v_vv_\rho}} \bigg({\Lambda^{d-3}I_1\over D_T^2} -{2m(\ell)\Lambda^{d-5}I_2\over D_T^4}+ \order(m^2)\bigg)\bigg]d\ell
      \,.\nn
       \label{mrr1}\\
\end{align}
\ew
Subtracting $m(\ell)$ from both sides of this expression and dividing by $d\ell$ gives us the differential recursion relation
\bew
\begin{align}
     {dm\over d\ell} &=(\chi_\rho-\chi+z-1)m+ {3 K_{d-1}u \rho_c^3\Delta\over \sqrt{v_vv_\rho}} \bigg({\Lambda^{d-3}I_1\over D_T^2} -{2m\Lambda^{d-5}I_2\over D_T^4}\bigg)+ \order(m^2)
      \,.\nn
       \label{mrr}\\
\end{align}
\ew
We now turn to the graphical corrections to $u$, which arise from the graph shown in figure \ref{uSquaredGraph}. This graph generates a correction to the right hand side of the spatially Fourier transformed equation of motion (\ref{TT v Perp Full}) for $v_i$ of the form
\bew
    \begin{align}
       \delta\bigg[ \pp_tv_i\bigg] = -{\rm FT}\{18iu^2q_i\}&\int^>_{{\bf p}, \Omega}C_{\rho \rho}(\bp, \Omega) G_{\rho v}(-\bp, -\Omega)(ip_\perp)\}\delta\rho^3(\br,t) \,.
       \end{align}
       \ew

\begin{figure}
    \centering
    \includegraphics[width=0.75\linewidth]{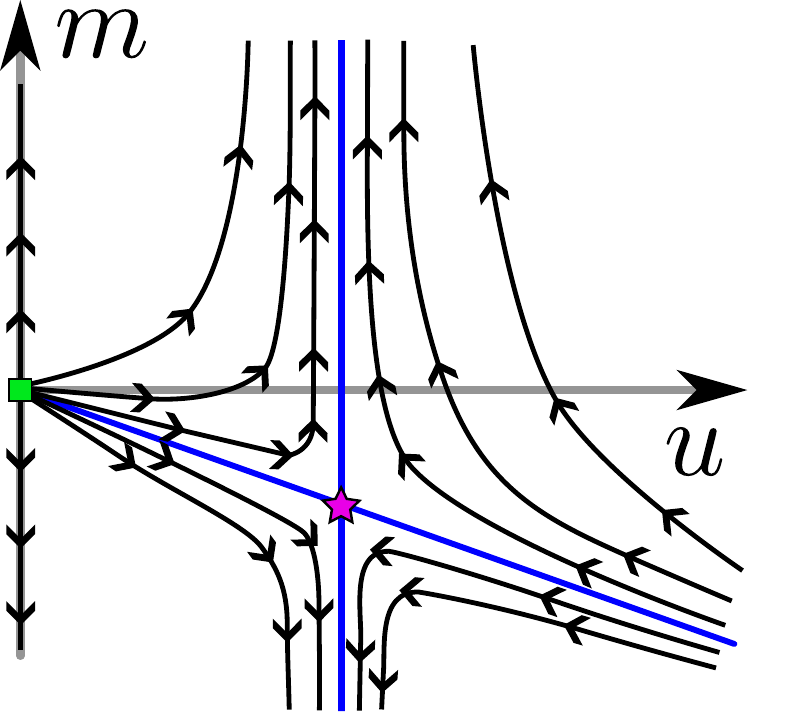}
    \caption{Renormalization group flows of the parameters $m$ and $u$ on the $u-m$ manifold of parameter space. The Gaussian fixed point is located at the origin and is marked by a green square. The linear fixed point is located a small $u$ and $m$ and is marked by a pink five pointed star.}
    \label{RG Flow Fig}
\end{figure}

   Because this is of precisely the same form as the $u$ term already present in the equation of motion \ref{TT v Perp Full}, we can identify this as a contribution $\delta[u]$ to $u$ given by    
       \begin{align}
       \delta[u] = 18u^2&\int^>_{{\bf p}, \Omega}C_{\rho \rho}(\bp, \Omega) G_{\rho v}(-\bp, -\Omega)(ip_\perp) \,.
       \label{ucorr}
       \end{align}
We can simplify this expression somewhat by noting, as we'll show {\it a posteriori}, that both $m$ and $u$ are of $O(\epsilon)$. Hence, this correction to $u$ is already of $O(\epsilon^2)$; keeping the dependence of $C_{\rho\rho}$ and $G_{\rho v}$ on $m$ will only add corrections of $O(\epsilon^3)$, which are negligible at the order to which we are working. A slightly more elaborate argument shows that keeping those $m$ terms in $C_{\rho\rho}$ and $G_{\rho v}$ only changes the critical exponents to $O(\epsilon^2)$, which is again higher order than that to which we are working. Therefore, we can drop the dependence on $m$ of $C_{\rho\rho}$ and $G_{\rho v}$ in \rf{ucorr}. Doing so, and using our earlier expressions { \ref{Crho} and \ref{GrhoV}} for $C_{\rho\rho}$ and $G_{\rho v}$ gives

\beq       
       \delta[u] = -18u^2\rho_c^3\Delta\int^>_{{\bf p}, \Omega}\frac{p_\perp^4\mu(\bp, \Omega)}{(\mu^2(\bp, \Omega) + \kappa^2(\bp, \Omega))^2}
       \label{ucorr2}
    \eeq
The alert reader will recognize the integral on the right hand side as that which already appeared in our expression { (\ref{Expanded B integral integrand})} for the linear in $m$ term in the renormalization of $m$. Indeed, making the change of variables { (\ref{rescale Omega px})}, we quickly find
\beq       
       \delta[u] = -{18u^2\rho_c^3K_{d-1}\Lambda^{d-5}I_2\over\sqrt{v_vv_\rho}D_T^4}d\ell    \,. 
       \label{ucorr3}
       \eeq
Using this result for the "graphs" in the first equality of our general RG recursion relation
\rf{uEoM} for $u$, and converting that recursion relation into a differential equation as we did the recursion relation for $m$, we find
\bew
\begin{align}
     {du\over d\ell} &=(3\chi_\rho-\chi+z-1)u- u^2\left({18 K_{d-1} \rho_c^3\Delta\Lambda^{d-5}I_2\over \sqrt{v_vv_\rho}D_T^4}\right)
      \,.\nn
       \label{urr}\\
\end{align}
\ew
To completely close these recursion relations, we need to determine the rescaling exponents $\chi$. $\chi_\rho$, and $z$. These are, of course, arbitrary. However, to make the analysis simple, it is very convenient to chose those exponents so as to keep the parameters of the linear theory (aside from $m$) fixed, because then we can treat the unknown integrals $I_1$ and $I_2$ as constants. To determine the choice of those exponents that accomplishes this we turn to the recursion relations { (\ref{lin rescale})} for the linear parameters, and note that, to one loop order, there are no graphical corrections to any of the linear parameters (aside from $m$). Hence, to one loop order, the choice of exponents that keeps those linear parameters fixed is the same as the choice we made in the previous subsection, in which we ignored those graphical corrections altogether; that is, 
\begin{align}
    z &= \zeta = 2 \,,
    \label{zlin}\\
    \chi &= \frac{1-d}{2} \,,
    \label{chilin}
  \\
    \chi_\rho &= \frac{3 - d}{2}   \,.
    \label{chirholin}
\end{align}
With these choices, our recursion relations for $m$ and $u$ become

\beqn
 {dm\over d\ell} &=&2m+ {3 K_{d-1}u \rho_c^3\Delta\over \sqrt{v_vv_\rho}} \bigg({\Lambda^{d-3}I_1\over D_T^2} -{2m\Lambda^{d-5}I_2\over D_T^4}\bigg) \nn\\\\
 {du\over d\ell} &=&\epsilon u- u^2\left({18 K_{d-1} \rho_c^3\Delta\Lambda^{d-5}I_2\over \sqrt{v_vv_\rho}D_T^4}\right) \,,
\label{rrs fin}
\eeqn
where we remind the reader that we've defined $\epsilon\equiv5-d$.
The flows implied by these equations, which are illustrated in figure { (\ref{RG Flow Fig})},  are extremely similar in {\it form} to those for equilibrium phase separation (that is, the equilibrium Ising model), but with $\epsilon$ being $5-d$ for our problem, while it is $4-d$ for the equilibrium problem.

The fixed points of these recursion relations can be determined by setting $ {dm\over d\ell}=0= {du\over d\ell}$, which leads to two equations for the two unknowns $(m^*, u^*)$, where the superscript $*$ denotes the fixed point values. Aside from the trivial Gaussian fixed point  $(m^*, u^*)=(0,0)$, which is doubly unstable for $\epsilon>0$ - that is, for $d<5$, and hence does not control the transition, there is a singly stable analog of the familiar Wilson-Fisher fixed point at
\begin{align}
m^*&=-\left({\Lambda^2D_T^2I_1\over12I_2}\right)\epsilon \,,\label{m*}\\
    u^* &= \left(\frac{ \sqrt{v_vv_\rho}D_T^4\Lambda^{5-d} }{18\rho_c^3\Delta K_{d-1} I_2}\right)\epsilon \label{u*}\,.
\end{align}
This is the "critical fixed point", which controls the transition  for $d<5$.
Linearizing the recursion relations
\rf{rrs fin} around this fixed point by taking
\begin{align}
    m(\ell) &= m^* + \delta m(\ell) \,,\\
        u(\ell) &= u^* + \delta u(\ell)
\end{align}

gives
\begin{align}
    \frac{d\delta m}{d\ell} &= (2 - {6K_{d-1}\rc^3\Delta\Lambda^{d-5}I_2\over\sqrt{v_vv_\rho}D_T^4}u^*)\delta m + C_u\delta u \,,\\
    \frac{d\delta u}{d\ell} &= (\epsilon-\left({36 K_{d-1} \rho_c^3\Delta\Lambda^{d-5}I_2\over \sqrt{v_vv_\rho}D_T^4}\right)\delta u \,.
\end{align}
where the constant 
\beq
C_u= {3 K_{d-1}u \rho_c^3\Delta\Lambda^{d-3}I_1\over \sqrt{v_vv_\rho}D_T^2}+O(\epsilon) \,.
\label{cu}
\eeq

Inserting the fixed point value \rf{m*} of $u$ into these recursion relations, we see that massive cancellations occur; in particular, the unevaluated integral $I_2$ cancels out. This leaves the recursion relations in the very simple form:
\begin{align}
    \frac{d\delta m}{d\ell} &= (2 - \frac{1}{3}\epsilon)\delta m + C_u\delta u\,,\label{linrec1}\\
    \frac{d\delta u}{d\ell} &= -\epsilon\delta u\,.\label{linrec2}
\end{align}
Seeking solutions to this linearized system of an exponential form; i.e., 
\begin{align}
    \begin{pmatrix}
        \delta m(\ell) \\
        \delta u(\ell)
    \end{pmatrix}
    =
   {\bf S} e^{\lambda\ell}
\end{align}
where {\bf S} is a constant eigenvector, and $\lambda$ a constant growth rate (which should not be confused with the parameter $\lambda$ in our original equations of motion!),  we see that there are two eigenvalues for $\lambda$:
\begin{align}
    \lambda_1 &= -\epsilon\\
    \lambda_2 &\equiv\lambda_t= 2 - \frac{1}{3}\epsilon
    \label{rgevs}
\end{align}
We identify $\lambda_t$, which is the only positive eigenvalue,  as the ``thermal" eigenvalue. We use this term in the usual RG sense, which is that it determines the dependence of the correlation length on the departure of the control parameter (which is usually temperature in equilibrium problems, hence the term "thermal eigenvalue").

We can  see this by the following completely standard RG analysis: 

Note that, as usual for a system of coupled linear ODE's, the general solution of our linearized recursion relations is
\begin{align}
    \begin{pmatrix}
        \delta u(\ell) \\
        \delta m(\ell)
    \end{pmatrix}
    =
    a(m_0) {\bf \hat{s}}_1 e^{-\epsilon\ell}+ c(m_0) {\bf \hat{s}}_2 e^{\lambda_t\ell} \,,
    \label{linsolgen}
\end{align}
where ${\bf \hat{s}}_{1,2}$ are the unit eigenvectors associated with the two eigenvalues $\lambda_{1,2}$ respectively, and the scalar constants 
   $a(m_0)$ and $c(m_0)$ are determined by the initial (i.e., bare) parameters of the model, and hence, in particular, the bare value $m_0$ of $m$.

The unit eigenvectors  are given to leading order in $\epsilon$ by
by:
\begin{align}
    s_1 &= {1\over\sqrt{4+C_u^2}}\begin{pmatrix}
           -C_u\\
            2
    \end{pmatrix},\\
    s_2 &= \begin{pmatrix}
        0\\
        1
    \end{pmatrix} \,.
\end{align}

These eigenvectors are depicted in figure \ref{RG Flow Fig} as the blue lines emanating from the linear fixed point.

Now it is easy to see that, in order for the initial parameters of our system to be such that the system flows under renormalization into the critical fixed point { (\ref{m*}) and (\ref{u*})}, we clearly must have
$\delta m(\ell\to\infty)\to0$,  $\delta u(\ell\to\infty)\to0$, which can clearly only occur at a value of $m_0$ such that $c(m_0)=0$, so that the exponentially growing part of the general solution { (\ref{linsolgen})} vanishes. We'll call this special value of $m_0$ $m_c$, that is, we'll define $m_c$ by the condition $c(m_c)=0$. It clearly then follows by analyticity that, for $m_0$ near v$m_c$, 
\beq
c(m_0)\approx A(m_0-m_c) \,,
\label{cexp}
\eeq
 with $A$ a constant. This implies that, if $A>0$, $c(m_0)$ will be positive for $m_0>m_c$. As a result, as RG time $\ell\to\infty$, the renormalized $m(\ell)$ will go to large, positive values (since the exponentially growing $c$ term in { (\ref{linsolgen})} will inevitably eventually dominate over the exponentially decaying $a$ term. This clearly corresponds to the {\it non}-phase separated region, in which the state of uniform density is stable. Likewise, if $m_0<m_c$, (continuing for now to assume $A>0$), $c(m_0)$ will be negative. As a result, as RG time $\ell\to\infty$, the renormalized $m(\ell)$ will go to large, negative values. This clearly corresponds to the phase separated region, in which the state of uniform density is unstable.

Hence, $m_c$ clearly corresponds to the critical value of $m$ at which the critical point  occurs.

Recognizing this, we can now use the expansion \rf{cexp} to obtain the behavior of the correlation length near the critical point by the following very standard RG argument

Starting with any $m_0$ near $m_c$, we run the renormalization group until we reach a value $\ell^*(m_0)$ of $\ell$ at which the renormalized $\delta m(\ell^*)$ takes on some particular "$O(1)$" reference value which we'll call $\delta m_r$. By "$O(1)$", which is literally meaningless in this context, since $\delta m$ is a dimensionful quantity, what we really mean is a value small enough that the linearized recursion relations { (\ref{linrec1}) and (\ref{linrec2})} , and, therefore, their solution { (\ref{linsolgen})}, remain valid, but as big as it can be consistent with that requirement. 

For $m_0$ close to $m_c$, therefore, the value of $\ell^*$ required to reach such a large $\delta m(\ell^*)$ will clearly be large, since the coefficient $c(m_0)$ of the exponentially growing part of the solution { (\ref{linsolgen})} of our linearized recursion relations is small in that case. It follows that, by the time $\ell$ reaches $\ell^*$, the exponentially decaying $a$ term in the solution { (\ref{linsolgen})} will have become negligible, so that 
\beq
\delta m(\ell^*)\approx c(m_0)e^{\lambda_t\ell^*}=\delta m_r \,,
\label{ell* cond}
\eeq
where in the last equality we have applied our condition on $\ell^*$ that it make $\delta m(\ell^*)=\delta m_r$.
It is clearly straightforward to solve \rf{ell* cond} for $\ell^*$; we'll instead solve it for $e^{\ell^*}$, which, as we'll see in a moment, proves to be the more useful quantity:
\beq
e^{\ell^*}=\left({\delta m_r\over c(m_0)}\right)^\nu\propto |m_0-m_c|^{-\nu_\perp} \,,
\label{ell*sol}
\eeq
where we've defined the "correlation length exponent"
\begin{align}
    \nu_\perp &= \frac{1}{\lambda_2}= \frac{1}{2} + \frac{\epsilon}{12} + \order(\epsilon^2)  \,.  
    \label{nuperpeps} 
\end{align}
Note that, in the limit $m_c\to m_0$, all of our starting systems have been mapped onto the same point
\begin{align}
    \begin{pmatrix}
        \delta u(\ell^*) \\
        \delta m(\ell^*)
    \end{pmatrix}
    =
   \begin{pmatrix}
        0\\
        \delta m_r
    \end{pmatrix} \,,
    \label{linsolgen}
\end{align}
since the exponentially decaying part of the solution { (\ref{linsolgen})} will have vanished in this limit.
Hence, all of these systems are mapped onto the same model, and, hence, onto a model with the same correlation lengths in the perpendicular and parallel direction.

Note that this does {\it not} imply that all of these systems have the same correlation lengths. On the contrary, since each of them will have to have been renormalized for a different, strongly $m_0-m_c$-dependent RG time as implied by { (\ref{ell*sol})}, they will have very different correlation lengths. Indeed, since, on every time step, we rescale lengths in the $\perp$-directions by a factor of $b=1+d\ell$, while directions in the $\parallel$-direction are rescaled (at the one loop order to which we've worked here) by a factor of $b^2$, the actual correlation lengths in the $\perp$  and  $\parallel$ directions $\xi_{\parallel,\perp}$ are related to those at the "reference point 
$ \begin{pmatrix}
        \delta u(\ell^*) \\
        \delta m(\ell^*)
    \end{pmatrix}
$
by
\beqn
\xi_{\perp}(m_0)&=&b^{n^*}\xi_{\perp}(\delta m_r)=e^{\ell^*}\xi_{\perp}(\delta m_r)
\propto|m_0-m_C|^{-\nu_\perp} \,, \nn\\\\
\xi_{\parallel}(m_0)&=&b^{2n^*}\xi_{\parallel}(\delta m_r)=e^{\ell^*}\xi_{\parallel}(\delta m_r)\propto|m_0-m_C|^{-2\nu_\perp}\nn\\&\equiv&|m_0-m_C|^{-\nu_\parallel}\,.\nn\\
\label{xis}
\eeqn
Here $n^*=\ell^*/d\ell$ is the number of RG steps required to reach $\ell^*$, and we've used the fact that, for $d\ell$ differential and $\ell^*=nd\ell$ finite, $b^{n^*}=(1+d\ell^*)^{n^*}
=(e^{d\ell^*})^{n^*}=e^{n^*d\ell^*}=e^{\ell^*}$. We've also used our earlier expression { (\ref{ell*sol})} for $e^{\ell^*}$, and defined 
\beq
\nu_{\parallel}=2\nu_\perp \,.
\label{nupardef}
\eeq

Since everything in our expressions \rf{xis} for these two correlation lengths is independent of the bare $m_0$ except for the terms explicitly displayed, which arise from the singular $m_0$-dependence of the RG time $\ell^*$, that explicitly displayed dependence is the entire dependence of the correlation lengths on $m_0$. Thus we have
\beqn
\xi_{\perp}(m_0)&\propto|m_0-m_C|^{-\nu_\perp} \sep
\xi_{\parallel}(m_0)\propto|m_0-m_C|^{-\nu_\parallel}\,,\nn\\
\label{xissum}
\eeqn
with the universal critical exponents $\nu_{\perp,\parallel}$ given by
\begin{align}
    \nu_\perp &= \frac{1}{\lambda_2}= \frac{1}{2} + \frac{\epsilon}{12} + \order(\epsilon^2) \sep  \nu_\parallel &= 1 + \frac{\epsilon}{6} + \order(\epsilon^2)  \,.  
    \label{nuperpeps} 
\end{align}

Note that the $2$ to $1$ ratio of the exponents $\nu_\parallel$ to $\nu_\perp$ will not persist to higher order in $\epsilon$; it is an artifact of the fact that the anisotropy exponent $\zeta=2$ up to and including linear order in $\epsilon$. At $O(\epsilon^2)$ and higher, there will be graphical corrections to, e.g., $D_{L\perp,\rho\perp}$ and various other parameters of the linear model, which will in turn make $\zeta\ne2$. The general {\it form} of \rf{xissum} will continue to hold at higher order in $\epsilon$, but the relation between $\nu_\parallel$ and $\nu_\perp$ will become
\beq
\nu_\parallel=\zeta\nu_\perp
\label{gen nu rel}
\eeq
with $\zeta$ given by
\beq
\zeta=2+O(\epsilon^2) \,.
\label{zeta eps}
\eeq

 We can obtain the critical exponent $\beta$, which gives the shape of the phase boundary in figure { (\ref{spinodal})}, by a similar argument. Because we rescale density by a factor of $b^{\chi_\rho}$ on each RG time step, the density difference $\delta\rho$ between the two coexisting densities in figure { (\ref{spinodal})} at some $m_0<m_c$ can be related to the corresponding difference 
at a {\it negative} reference value $\delta m_r$ via
\begin{align}
    \Delta\rho(\delta m_0; \ell =0) &= e^{\ell_* \chi_\rho}\Delta\rho(\delta m_r) \,.\\
    \label{TIrho}
\end{align}
As we argued for the correlation length, so here the quantity $\Delta\rho(\delta m_r) $ will be independent of our starting value $\delta m_0$. Hence, all of the dependence of 
$\Delta\rho$ on $m_0$ comes again from the factor $e^{\ell_* \chi_\rho}=\left(e^{\ell_*}\right)^ {\chi_\rho}$. The relation { (\ref{ell*sol})} relating $e^{\ell_*}$ to $m_0$ continues to hold, so we have

\begin{align}
    \delta\rho &\propto |m_0-m_c|^{\beta}
    \label{beta}
\end{align}
with
\beq
\beta=-\nu_\perp\chi_\rho \,.
\label{betanuchi}
\eeq
This relation is exact. We can use it to generate an $\epsilon$-expansion for $\beta$ by using our earlier expansion { (\ref{nuperpeps})} for $\nu_\perp$, along with our earlier argument that $\chi_\rho$ is unchanged from its linear value \rf{chirholin} to first order in $\epsilon$, in \rf{betanuchi} to obtain
\beqn
\beta&=&-\nu\chi_\rho=\left(\frac{1}{2} + \frac{\epsilon}{12} + \order(\epsilon^2)\right)\left({3-d\over2}+O(\epsilon^2)\right)\nn\\&=&\left(\frac{1}{2} + \frac{\epsilon}{12} + \order(\epsilon^2)\right)\left(-1+{\epsilon\over2}+O(\epsilon^2)\right)\nn\\&=&\frac{1}{2} - \frac{\epsilon}{6} + \order(\epsilon^2) \,.\nn\\
\label{betaeps}
\eeqn

\section{discussion and summary}\label{con}

We have shown that phase separation in polar ordered active fluids ("flocks" )belongs to a new universality class, radically different from that of equilibrium phase separation. Even the critical dimension of the problem is different: flocks have a critical dimension of five, while for equilibrium phase separation, the critical dimension is four.

The well-informed reader will have noticed that, to leading order in $\epsilon$, our results for the thermal eigenvalue $\lambda_t$ and the order parameter exponent $\beta$ are identical, {\it when expressed in terms of} $\epsilon=d_c-d$, to those for the equilibrium problem. The only difference, {\it at this order}, is that $d_c$ is larger by $1$ for the flocking problem that in equilibrium.

One might therefore bet tempted to speculate that this property holds to all orders; that is, that $\lambda_t^{\rm flock}(d)=\lambda_t^{\rm equil}(d-1)$. However, such a speculation is almost certainly incorrect. 
There is, in fact, at least one known problem in which, as in ours, the critical dimension differs from that for equilibrium phase separation by $1$, and the exponents, to leading order in $\epsilon$, are the same as those of the equilibrium short-ranged Ising model. That problem is the Ising model with dipolar interactions\cite{Idip1, Idip2}, for which $d_c=3$. For that problem, a second order in $\epsilon$ calculation\cite{Idip2} finds different results  from the second order in $\epsilon$ calculation for the short-ranged problem. This problem has many features in common with ours, including a two to one anisotropy of scaling at the linear level, and the fact that correlations do not decay exponentially even beyond the correlation length. 
Hence, we strongly suspect that our problem does {\it not} map on to the equilibrium problem in one lower dimension.

Our $\epsilon$-expansion results are, of course, {\it not }expected to be accurate for the 
physically relevant cases of $d=3$ ($\epsilon=2$) or $d=2$ ($\epsilon=3$). The significance 
of our calculation, and, in particular, our demonstration of the existence of a non-Gaussian 
fixed point in $d=5-\epsilon$, is that he shows the existence of a critical point in a new 
universality class for phase separation in flocks.

 What can we say about the exponents in $d=3$ and $d=2$? One thing that we know is that, at higher loop order, linear coefficients like the diffusion constants $D_{L\perp}$ and so on will get graphical corrections. This will make the anisotropy exponent $\zeta$ differ from $2$, although it will remain universal. Thus we will continue to have two correlation lengths, one along ($\parallel$). and one perpendicular to ($\perp$) the direction of mean flock motion. These will diverge as the control parameter $m_0$, defined via the pressure expansion { (\ref{Pexp})}, approaches its critical value $m_c$, according to 
\beqn
\xi_{\perp}(m_0)&\propto|m_0-m_C|^{-\nu_\perp} \sep
\xi_{\parallel}(m_0)\propto|m_0-m_C|^{-\nu_\parallel}\,,\nn\\
\label{xissumconc}
\eeqn
with the universal critical exponents $\nu_{\perp,\parallel}$ obeying
\begin{align}
\nu_\parallel = \zeta \nu_\perp  \,, 
    \label{nuperpepsconc} 
\end{align}
with
\beq
\zeta=2+O(\epsilon^2) \,.
\label{zeta eps}
\eeq
This $\epsilon$-expansion obviously tells us nothing quantiative about the universal anisotropy exponent $\zeta$ in $d=2$ or $d=3$, other than that it does not equal $2$.

One very big remaining open question is: what is the {\it lower} critical dimension 
$d_{LC}$,  below which the critical point disappears, for our problem? For equilibrium phase separation, that lower critical dimension is one. The "$d\to d-1$"
argument described above would then suggest that $d_{LC}=2$ for our problem. But, as discussed earlier, we do not believe that argument, so the question of the lower critical dimension remains open. Thus, we cannot unambiguously claim that our new universality class persists down to the physically relevant spatial dimensions $d=2,3$. This is a question that will have to be answered by experiment. What we can be quite confident of in light of our work is that, if such a critical point {\it does} exist, it will certainly belong to a universality class very different from that of equilibrium phase separation.

\bibliographystyle{apsrev4-1}
\end{document}